\documentclass{ws-ijmpa}
\usepackage[super,compress]{cite}
\usepackage{graphicx}
\begin{document}
\markboth{George W.-S. Hou}{HEP Window on the Universe}

%
\catchline{}{}{}{}{}
%

\title{
Perspectives and Outlook from HEP Window on the Universe
}

\author{George Wei-Shu Hou
}

\address{
Department of Physics, National Taiwan University\\
Taipei 10617, Taiwan\\
wshou@phys.ntu.edu.tw
}

\maketitle

\begin{history}
\received{Day Month Year}
\revised{Day Month Year}
\end{history}

\begin{abstract}
This brief review grew out from the HEP Concluding Talk of 
the {\it 25th Anniversary of the Rencontres du Vietnam}, held August 2018 in Quy Nhon.
The first two-thirds gives a Summary and Highlights,
or snapshot, of High Energy Physics at the end of Large Hadron Collider (LHC) Run 2.
It can be view as the combined effort of the program organizers, 
the invited plenary speakers, and finally filtered into the present mosaic.
It certainly should not be viewed as comprehensive.
In the second one-third, a more personal Perspective and Outlook is given,
including my take on the flavor anomalies, and why the next 3 years, the
period of Long Shutdown 2 plus first year (or more) of LHC Run~3,
would be {\it Bright} and {\it Flavorful}, 
with much hope for uncovering {\it New Physics}.
We advocate extra Yukawa couplings as {\it the most likely, next, New Physics}
to be tested, the effect of which is already written in
our Matter Universe.

\end{abstract}

\ccode{PACS numbers:}


\section{HEP Summary and Highlights at End of LHC Run 2}

\vskip0.06cm
\centerline{\it
 Happy 25th Anniversary,
   Rencontres du Vietnam!!
}
\vskip0.2cm
I returned to Taiwan in 1992 and missed the first {\it Rencontres du Vietnam} held in 1993.
But, lured by the 1995 total solar eclipse, I did join the second one held in Ho Chi Minh City. 
I still have the conference T-shirt, which I wore as a personal tribute when giving 
the ``HEP Summary and Outlook'' at the {\it 25th Anniversary}.\cite{RduVweb}

In the first part of this brief review, which is adapted from the talk, 
I will cover the Highlights at the LHC and the corresponding experimental subjects, 
then flavor physics,\footnote{
The main discussion of the ``flavor anomalies'' are deferred to the Outlook part.
} 
neutrinos and Dark Matter (DM), then
theory, and finally on Asia in the World.
As there were only 7 theory talks out of a total of 37 plenary talks 
(I apologize for not covering parallel talks,
 as well as any other negligence), our emphasis would be on experiment.
In the second part, I will offer a Perspective and Outlook
on the {\it HEP Window on the Universe}.

Our emphasis would be the HEP physicists' yearning for New Physics,
i.e. physics Beyond the Standard Model (BSM).

\begin{figure}[t]
\centerline{\includegraphics[width=7cm]{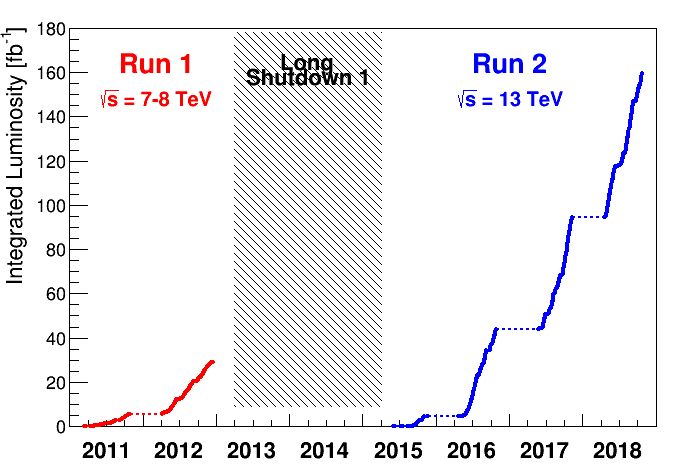}}
\caption{
 Integrated proton-proton luminosity for LHC Run 1 (7 and 8 TeV) and Run 2 (13 TeV).
 [Source: https://twiki.cern.ch/twiki/bin/viewauth/LhcMachine/LhcCoordinationMain]. \label{intLumiLHC}}
\end{figure}

\subsection{LHC as Window on the Universe}

The LHC has delivered spectacular performance at Run 2 (see Fig.~\ref{intLumiLHC}),
where a total of  $\sim 150$ fb$^{-1}$ pp collision data were 
recorded by both ATLAS and CMS before entering Long Shutdown~2
  (LS2, namely 2019--2020), while LHCb collected about
twice the data of Run 1, i.e. $\sim 6$ fb$^{-1}$,
but at higher $b\bar b$ cross section.

The ALICE experiment, which celebrated its own\cite{Nayak} 25th anniversary
(ATLAS and CMS celebrated a year earlier),
asks for special Pb-Pb, Xe-Xe, p-Pb,\footnote{
 The p-Pb and Xe-Xe collisions were not in the original design for heavy ion runs!
} 
as well as pp collision runs at various energies,
and epitomizes the LHC as a {\it Window on the Universe}:
if Astrophysics, as discussed in our companion {\it Window on the Universe}
 conference,\cite{RduVweb} takes us back to 380,000 years after the Big Bang,
 such as with CMB (Cosmic Microwave Background),
then the study of quark-hadron phase transitions and 
exploring the quark-gluon plasma takes us back to 
within a few microseconds after the Big Bang.
Experimental topics cover\cite{Nayak} strangeness enhancement, resonance scattering
and nuclear modification effects,
with energy densities at LHC reaching several times that of RHIC,
the predecessor at Brookhaven.
One probes timescales and (hydro)dynamics, 
strongly coupled liquid with small viscosity, phase diagram, etc.,~\cite{Varma}
which are relevant to the Early Universe.

Of course, the Energy and Intensity (as well as neutrino) frontiers covered
below all offer Windows on our Universe, pointing towards much earlier times.

\subsection{SM and BSM Higgs}

Since its observation in 2012,\cite{ATL-h,CMS-h} 
the 125 GeV boson has been demonstrated to resemble 
the Standard Model (SM) Higgs boson\cite{Cepeda,Khachatryan:2016vau} 
remarkably well!

One major highlight of 2018 is the completion of {\it direct} measurements, 
by both ATLAS and CMS, of third generation Yukawa couplings:
all were found to be consistent with SM expectations.
This started with the jet-assisted observation (when combined with Run 1)
of $H \to \tau^+\tau^-$ in 2017 by CMS,\cite{Sirunyan:2017khh}
and the subsequent observation\cite{Sirunyan:2018cpi} using Run 2 data alone.
Then there was the observation\cite{Sirunyan:2018hoz,Aaboud:2018urx} 
of $t\bar tH$ production before summer 2018 (more on this later in the Outlook part).
And finally, the observation of $H \to b\bar b$ in $VH$ production,
which was officially announced in a joint seminar at CERN in late August 2018,
shortly after the {\it Rencontres du Vietnam}.
ATLAS had already announced it at ICHEP2018 in Seoul,\cite{TCarli}
but the CMS Spokesperson traveled to Quy Nhon for
the make-up first announcement,~\cite{Butler} culminating in the two subsequently 
submitted, and readily accepted, papers.\cite{Hbb-ATL,Hbb-CMS}
What is remarkable is that $Z\to b\bar b$ provides ``{\it in situ}'' validation of the method,
where excess above the $Z$ can be clearly seen, and found consistent
with $H \to b\bar b$ in mass and cross section.
Combining all available data, ATLAS\cite{Hbb-ATL} and CMS\cite{Hbb-CMS} 
give the $H \to b\bar b$ signal strength
$\mu = 1.01 \pm 0.20$ and $1.04 \pm 0.20$, respectively,
for the ratio compared with SM expectation. 
Note that both central values are remarkably close to 1,
although each were a combination of several different processes.

\begin{table}[t]
\tbl{Signal strength $\mu$ for Higgs boson coupled to third generation fermions.
   See text for discussion and references.}
{
\begin{tabular}{@{}cccc@{}} \toprule
             & $\mu_{{\rm H}\tau\tau}$ & $\mu_{\rm ttH}$ & $\mu_{\rm Hbb}$ \\
 \colrule
ATLAS & $1.09^{+0.35}_{-0.30}$ & \hphantom{0} $1.32^{+0.28}_{-0.26}$ & $1.01 \pm 0.20$ \\
CMS    & $1.24^{+0.29}_{-0.27}$ & \hphantom{0}  $1.26^{+0.31}_{-0.26}$ & $1.04 \pm 0.20$ \\
 \botrule
\end{tabular} \label{ta1}
}
\end{table}

The measured signal strength, or $\mu$, values are given in Table~\ref{ta1}.
Note that for ATLAS, it is combining with Run 1 that\cite{Aaboud:2018pen}
turns $H \to \tau^+\tau^-$ into an observation.
With the Yukawa couplings $\lambda_\tau$, $\lambda_b$ and $\lambda_t$
all directly measured and found consistent with SM,\footnote{
 We note in passing that CMS has probed the sign of top Yukawa coupling
 via $tH$ production,\cite{Sirunyan:2018lzm} favoring the SM positive sign.
} 
the drive\cite{Cepeda} now is for the 13 TeV combined fit, 
for differential distributions, $H \to \mu^+\mu^-$, and di-Higgs (or $HH$) production.

For BSM Higgs bosons,\cite{Ilic} one typically adds
an extra scalar singlet (real or complex), doublet (2HDM), or triplet, or some combination.
2HDM is the most popular form, with usual notation
of $H^0$, $A^0$ ($CP$-odd scalar) and $H^\pm$ for the exotic bosons,
while the observed 125 GeV boson is denoted as $h^0$.
Having an extra triplet offers a doubly-charged $H^{\pm\pm}$ boson,
but otherwise it is not so easy to distinguish from a 2HDM.
The most popular model is 2HDM II that is automatic in minimal SUSY (MSSM):
up-type quarks couple to one doublet, 
down-type quarks (and charged leptons) to the other doublet.
As alluded to already, a special phenomenon that emerged from LHC Run 1 is that
$\cos(\beta - \alpha)$, the $h^0$--$H^0$ mixing angle between 
$CP$-even Higgs bosons in 2HDM II,
appears to be small, i.e. $h^0$ is rather close to the SM Higgs boson.\cite{Khachatryan:2016vau}
This {\it alignment} phenomena, that Run 1 data 
prefers the  $\cos(\beta - \alpha) \to 0$ limit,
should be pursued with vigor using full Run 2 data.

BSM Higgs bosons are searched for in many channels by ATLAS and CMS,
exploiting interesting techniques in boosted analyses and 
background estimations with multivariate analysis (MVA).
Unfortunately, no significant excess has been found so far.
Sample searches~\cite{Ilic} with 13 TeV data include: 
\begin{itemlist}
 \item Charged Higgs $H^\pm$, e.g. in\cite{H+taunu-ATL,CMS:2018ect} $H^+ \to \tau^+ \nu_\tau$
 and\cite{H+tb-ATL} $H^+ \to t\bar b$ processes;
\item Neutral $H^0,\ A^0 \to \tau^+\tau^-$ with
 large $\tan\beta$ enhancement,\cite{S0tautau-CMS}
 as exclusion in $m_{A^0}$--$\tan\beta$ plane;
\item $h(125) \to a^0a^0 \to \mu^+\mu^-\tau^+\tau^-$,\cite{haa-mumutata-CMS} 
 where $a^0$ is a light exotic pseudoscalar;
\item  $\ell\nu q\bar q$ search\cite{lnuqq-ATL}  for heavy scalar that can 
decay to $WV$;\footnote{
 A mild local excess (less than 3$\sigma$) is found above 1.5 TeV.
}
and
\item $h(125) \to Z_dZ_d \to \ell\ell\ell'\ell'$,\cite{lll'l'-ATL} where $Z_d$ is a dark boson.
\end{itemlist}

\subsection{SUSY}

Absence of SUSY~\cite{Sonneveld} is the single most important
non-observation so far at the LHC.

\begin{figure}[b]
\centerline{\includegraphics[width=12.5cm]{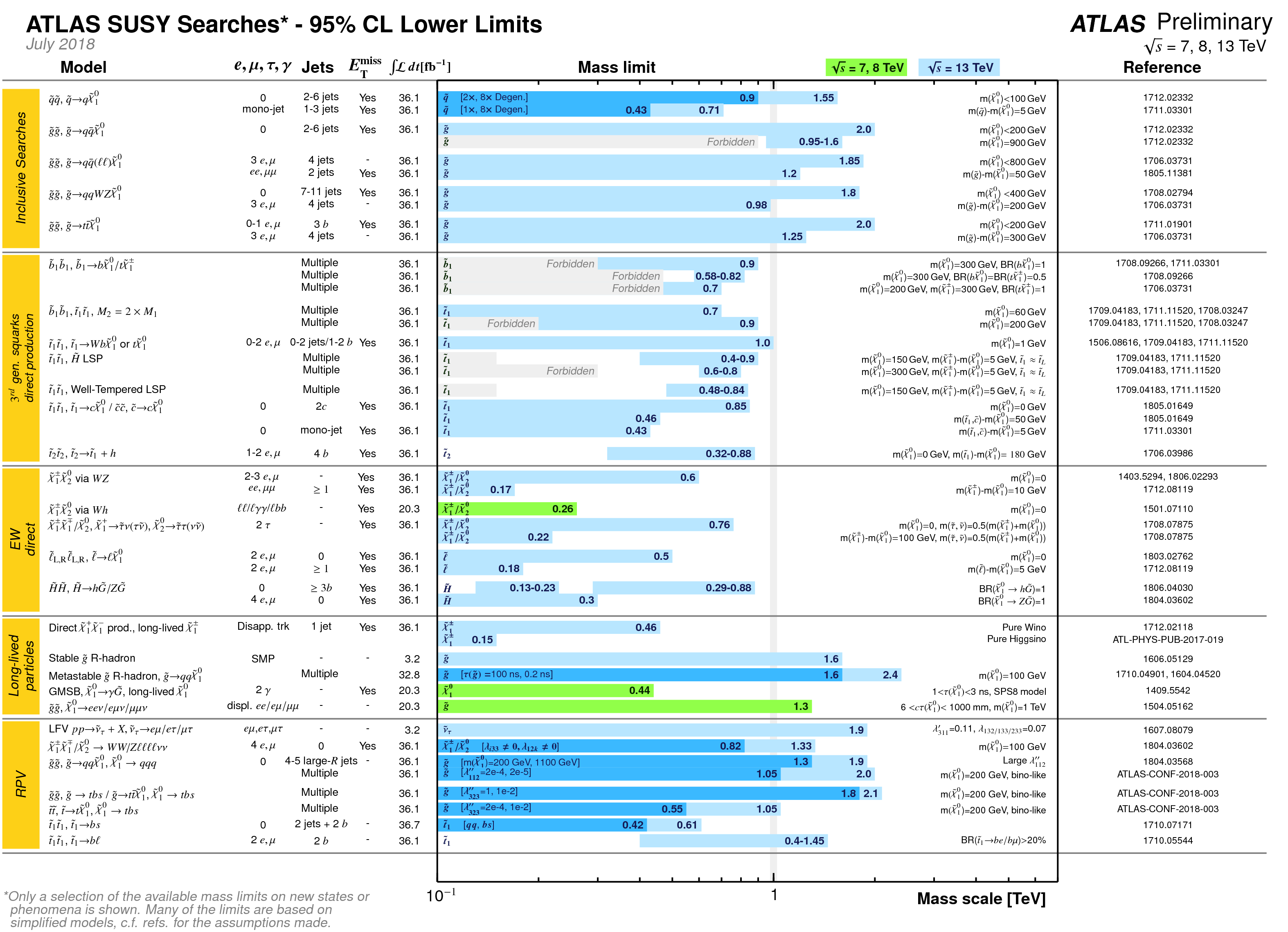}}
\caption{
 Summary table of ATLAS SUSY search lower limits at 95\% C.L.,
 as of July 2018
 [Source: https://atlas.web.cern.ch/Atlas/GROUPS/PHYSICS/CombinedSummaryPlots/SUSY/]. \label{SUSY18}}
\end{figure}

SUSY is a great theory that could\cite{Butler} (help) solve three problems at once:
the hierarchy problem, unification of gauge couplings, DM.
In 2010, many thought SUSY would be seen 
soon after startup, with just $100$ pb$^{-1}$:
 some expected it to be the first major LHC discovery,
 even before the Higgs.
But, as Run 2 has come to an end, the reality is that SUSY is so far a ``No Show''.
Perhaps\cite{Butler} it is heavier than we thought,
perhaps it is more devious or obscure, e.g. more weakly coupled,
or not fulfilling all three tasks.
Could it be R-Parity Violating (RPV), or hiding via {\it Long-Lived Particles} (LLP)?
Or maybe Nature simply did not adopt it at this scale.

Many good ideas are still being explored,~\cite{Butler,Sonneveld,Polini}
however, and SUSY is still a vibrant area of research, such as~\cite{Polini}
1)~compressed spectra (small mass splittings), 
2)~longer decay chains (less missing $p_T$), 
3)~lower rates (or~\cite{Butler} ``electroweakinos''), or complexities such as 
4)~LLPs (disappearing tracks, emerging jets), or 5)~RPV.
We will refer to these as 1)--5) in our Outlook,
which may also refer to the five classes given in Fig.~\ref{SUSY18},
where one can see the tremendous effort at ATLAS
(with corresponding counterpart at CMS) with 13 TeV data.
Although tendency is towards exploring higher masses with simplified models,
there is plenty of unexplored model space at low mass.

It is now the duty of experiments to leave no stones unturned.

\subsection{BSM/EXO}

We know something new must appear somewhere, but what is the scale?

With emphasis on {\it very heavy} BSM particles, searches~\cite{Pazzini} cover
 dibosons ($X \to WH$, $ZH$,  $VV'$ and $HH$,
  as we have already seen in BSM Higgs search); 
 vector-like quarks, or VLQ
  ($T \to bW$, $tZ$, $tH$, and $B \to tW$, $bZ$, $bH$), 
  motivated as ``top-partners'' to help alleviate the hierarchy problem; and 
 bosonic resonances with top in final state
  ($W' \to t\bar b$, $Z' \to t\bar t$, $t\bar T$, and
  third generation leptoquark ${\rm LQ}_3 \to \tau \bar t$).
%
These very massive particles produce highly boosted SM objects,
hence reconstruction and identification of {\it hadronic} decays
 become critical to most searches.
The hadronic decay tools often involve boosted jets or jet substructure
at high $p_T$, such as a $W$-, $Z$- or $H$-jet, or a top-jet that merges a $b$- and $W$-jet,
and MVA techniques are exploited to maximize the power of 
available statistics. In particular,  we have witnessed rapid growth
 since 2017 in the application of deep neural nets and machine learning.~\cite{Butler,Radovic:2018dip,Guest:2018yhq}

A common analysis issue is to tell a $W/Z \to q\bar q$ 
(or $t\to bq\bar q$, or $H \to b\bar b$) jet from a QCD jet,
and one typically goes through~\cite{Pazzini} 
\begin{itemlist}
 \item{Reconstruction}: for example, CMS uses Particle Flow\footnote{
 Particle Flow\cite{Sirunyan:2017ulk} uses all available information to reconstruct physics objects,
 and produces a big improvement in jet energy resolution, tau-lepton identification,
 and helps with high pileup.
 It paves the way for future data analysis, and even detector design,
 at high energy hadron colliders.
} 
candidates as starting objects, and mitigate high pileup
by using special tools such as PUPPI (PileUp Per Particle Identification);\cite{Bertolini:2014bba}
 \item{Grooming}: removal of soft and large-angle radiation to recover mass,
 such as (Jet) Trimming\cite{Krohn:2009th} used in ATLAS,
  i.e. removal of any subjet within a cone of $R = 0.2$ with less than 5\% of the jet $p_T$;
 \item{Tagging}: assign a ``tag'' to a jet, based on likelihood for ``signal'' or QCD,
 such as the use of Trimmed-jet mass, $N$-subjettiness,\cite{Thaler:2010tr}
 and the ratio of the energy correlation functions\cite{Larkoski:2014gra} $D_2$ by ATLAS.
\end{itemlist}

\begin{figure}[t]
\hskip-1.1cm{\includegraphics[width=13.7cm]{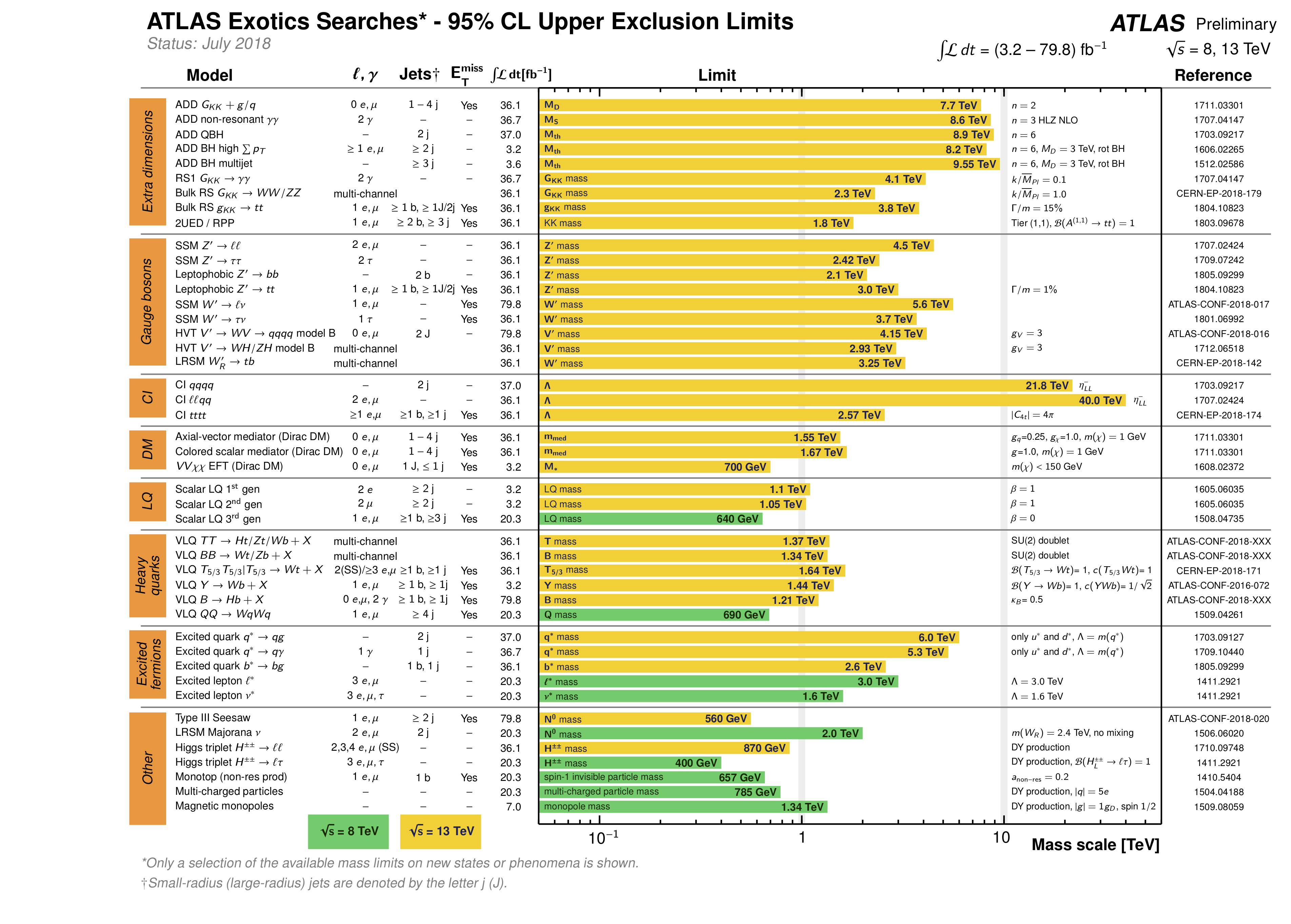}}
\caption{
 Summary table of ATLAS Exotics search lower limits at 95\% C.L.,
 as of July 2018
 [Source: https://atlas.web.cern.ch/Atlas/GROUPS/PHYSICS/CombinedSummaryPlots/EXOTICS/]. \label{Exotics18}}
\end{figure}

Sample search results in 2018, based on Run 2 data, are 
\begin{itemize}
 \item $X\to VV'$ search in all-hadronic final states, via two highly-energetic large radius jets
     (ATLAS, $\sim 80$ fb$^{-1}$);\cite{ATLAS:2018tpf}
 \item $X \to VH$ search with leptonic $W$ and $Z$ decays and $H\to b\bar b$ merged jets
     (CMS, $\sim 36$ fb$^{-1}$);\cite{Sirunyan:2018qob}
 \item $X \to WV$ search with $W \to \ell\nu$ plus $V \to q\bar q$ as single large radius jet
     (CMS, $\sim 36$ fb$^{-1}$);\cite{Sirunyan:2018iff}
 \item Search for VLQ pair production, namely $T\bar T$ or $B\bar B$ decay into
   final states with jets and no leptons,\cite{Aaboud:2018wxv}
   as well as a combination paper\cite{Aaboud:2018pii} of all VLQ pair production searches
     (ATLAS, $\sim 36$ fb$^{-1}$);
 \item Search for single VLQ production, i.e. $B\to tW$ with $t$ and $W$ highly boosted
     (CMS, $\sim 36$ fb$^{-1}$);\cite{Sirunyan:2018ncp}
 \item Search for $W' \to T\bar b$, $\bar B t$ in the fully boosted $tH\bar b$ final state, 
   involving $t$ and $H$ tagging besides usual $b$-tagging
     (CMS, $\sim 36$ fb$^{-1}$).\cite{Sirunyan:2018fki}
 \item Search for $Z' \to T\bar t$ production in lepton + jets, 
   with $T\to bW$, $tZ$ and $tH$, where one top in $tZ\bar t$ or $tH\bar t$
   decays semileptonically
     (CMS, $\sim 36$ fb$^{-1}$);\cite{Sirunyan:2018rfo}
\end{itemize}
The boosted jet approach is intrinsically more sensitive for higher masses. 
Thus, heavy bosons are probed up to 3--4 TeV, and
heavy fermions are probed up to 2 TeV.
There are no clear signals so far.

Besides heavy BSM particles, there is a variety of other
Exotics searches. In ATLAS, Exotics is defined as BSM without SUSY, 
though in CMS there is another group called B2G (Beyond 2nd Generation) 
aside from EXO, where EXO is the most productive physics group.
Any given presentation on BSM/EXO search can cover only some limited slice of
possible final states probed by ATLAS (see Fig.~\ref{Exotics18} for
a Summer 2018 summary list) and CMS, and even LHCb.
We give a snapshot of EXO topics~\cite{Milic} covered at {\it Rencontres du Vietnam}:
\begin{itemize}
 \item Dijet resonance $X$ (e.g. technipion, $Z'$) search with $\ell = e,\; \mu$ trigger,
    probing $m_{jj}$ down to 0.25 TeV and up to 6 TeV 
     (ATLAS, $\sim 80$ fb$^{-1}$);\cite{l+jj-ATL-CONF}
 \item Inclusive dark photon $A' \to \mu^+\mu^-$ search, both prompt and long-lived
     (LHCb, 1.6 fb$^{-1}$),\cite{Dphot-LHCb,Kouskoura}
    which is a demonstration study by LHCb that follows 
    a  novel and promising phenomenological proposal\cite{Ilten:2016tkc} targeting Run 3;\footnote{
 Interest in DM and dark photon search is not limited to LHC experiments.
 For instance, Belle has a recent result\cite{Seong:2018gut} that uses the dipion as tag in
 $\Upsilon(2S) \to \Upsilon(1S) \pi^+\pi^-$ decay,
 to search for $\Upsilon(1S) \to \gamma\chi\chi$, where $\chi$ is a low mass DM particle, 
 with $\chi\chi$ off-shell or in resonance (from an $A^0$ mediator).
 Many flavor or other experiments have pursued DM or dark particle searches
 in various ways that we have not been able to cover.
}

\item Search for heavy Majorana neutrino $N$ in same-sign dileptons plus at least one jet,
    i.e. $\ell N(q)$ production followed by $N \to \ell W$
     (CMS, $\sim 36$ fb$^{-1}$);\cite{nR-CMS}
\item  Search for Lepton-Flavor Violation (LFV): $X \to \ell\ell'$, where $X$ could be a $Z'$
    or $\tau$-sneutrino, hence $\ell$ includes $\tau$
     (ATLAS, $\sim 36$ fb$^{-1}$);~\cite{lfvZ'-ATL}
\item Search for stopped LLPs, or Displaced Vertices, in calorimeter or muon system,
    during period of $> 700$ hours well separated from collision data
     (CMS, $\sim 39$ fb$^{-1}$).~\cite{Butler,LLP-CMS}
\end{itemize}
To complement the last item, on theory side, a model of mirror fermions was presented,~\cite{Hung} 
where an electroweak scale $\nu_R^{\rm M}$ that belongs to a weak right-handed doublet 
offers~\cite{Chakdar:2016adj} an example for LLP signature.\footnote{ 
 As we have mentioned LLPs many times already, we note the
 MATHUSLA (MAssive Timing Hodoscope for Ultra-Stable neutraL pArticles)
 proposal,\cite{Chou:2016lxi} 
 a dedicated large-volume displaced vertex detector for the HL-LHC,
 which has now put forth the Letter of Intent.\cite{Alpigiani:2018fgd}
 To operate on the surface above ATLAS or CMS, it claims better sensitivity
 than the two experiments by several orders of magnitude,
 and can search for LLPs at GeV mass or higher, up to $c\tau \sim 10^7$ m.
} 
The model provides a test of seesaw mechanism and a solution to the strong CP problem.

\begin{figure}[ht]
\centerline{\includegraphics[width=12.5cm]{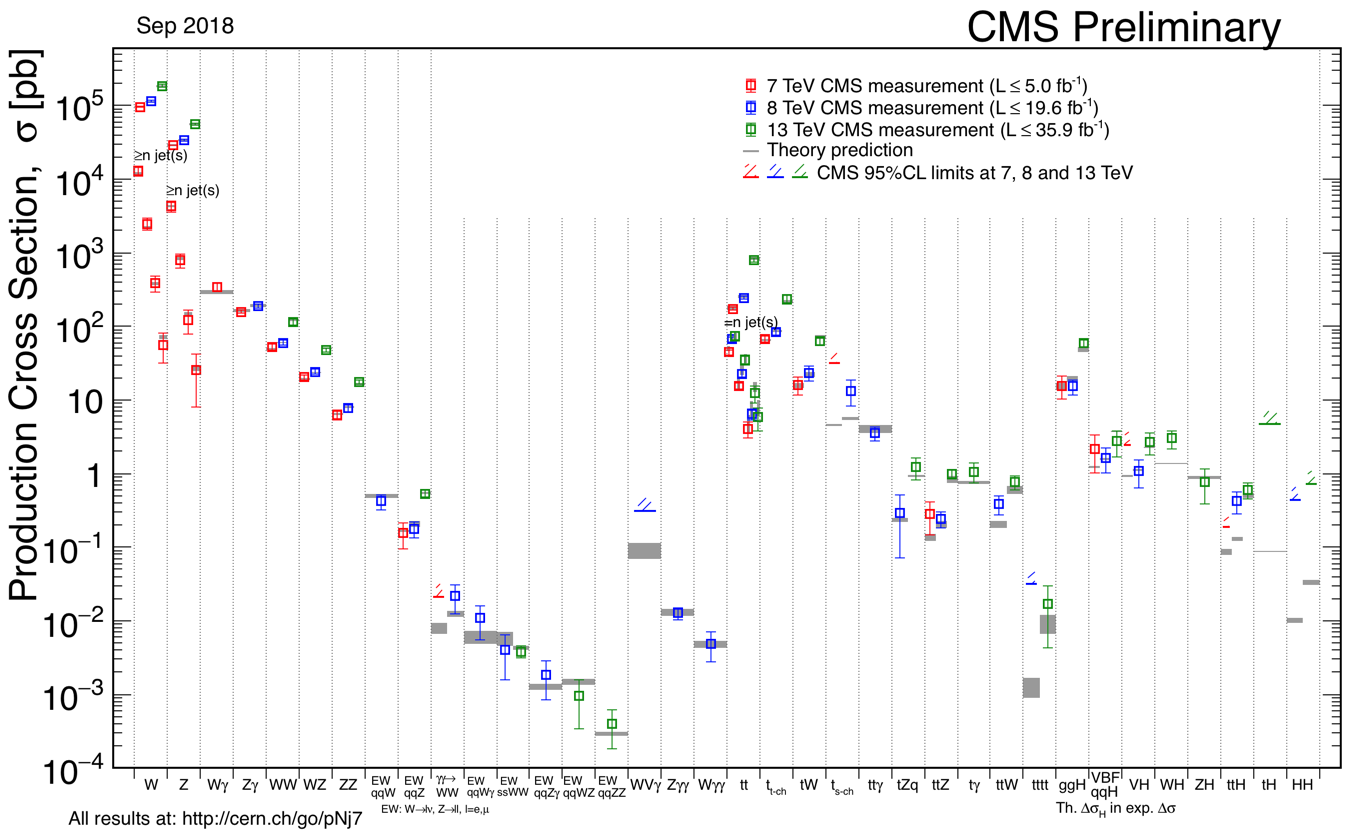}}
\caption{
 Summary of Standard Model cross section measurements by CMS as of September 2018
 [Source: https://twiki.cern.ch/twiki/bin/view/CMSPublic/PhysicsResultsCombined]. \label{SMP18}}
\end{figure}

\subsection{SM} 

Returning from Beyond SM back to SM itself,~\cite{Bella} all measurements confirm it
(see Fig.~\ref{SMP18}, which spans almost 10 orders of magnitude), 
but in the context of {\it Windows on the Universe}, 
SM measurements play the essential role in testing our current 
understanding of the {\it laws that govern} the Universe. 
On the other hand, for continued pursuit at High Energy and High Intensity frontiers,
SM processes gives {\it background} to all searches. 

Let us mention a few highlights:~\cite{Bella}
\begin{itemize}
 \item Weak mixing angle $\sin^2\theta_{\rm eff}^l$: 
    The ATLAS 8 TeV combined error~\cite{theta^l-ATL-CONF} of $\pm 0.00036$ is
    approaching LEP single experiment sensitivity, as well as the
    Tevatron combined error~\cite{Hirosky} at $\pm 0.00033$.
    This appears quite promising, and we look forward to the full Run 2 data update,
    which probably would take some time.
 \item $W$ mass:
   ATLAS 8 TeV measurement reaches an accuracy of $19$ MeV,~\cite{Wmass-ATL}
   which is already better than LEP, and approaching the Tevatron combined error.
\item Vector boson scattering (VBS) observations:
    CMS has observed,\cite{SS-WWjj-CMS} with $\sim 36$ fb$^{-1}$ at 13 TeV,
    VBS production of same-sign $W^\pm W^\pm$ at 5.5$\sigma$ (5.7$\sigma$ expected),
    and the measured fiducial cross section is in agreement with LO theory prediction.
    ATLAS has observed~\cite{SS-WWjj-ATL-CONF} the mode at 6.9$\sigma$
      (4.6$\sigma$ expected by Sherpa) based on data of similar size, while
    observing~\cite{WZjj-ATL-CONF} also the electroweak production of $W^\pm Z$
    plus two jets at 5.6$\sigma$ (3.3$\sigma$ expected). These two observations, however,
    are still at conference paper stage.
    In any case, since higher order calculations are lacking except\cite{Biedermann:2016yds}
    for same-sign $W^\pm W^\pm jj$,
    experiment is pushing theory to make progress on modes such as $WZjj$.
\end{itemize}

We see that the weak mixing angle and $W$ mass measurements at the LHC 
compare well with the GFitter global electroweak fit results.\cite{Haller:2018nnx}
In the longer term, VBS can check the unitarity of $VV \to VV$ scatterings,
to confirm the role of the Higgs boson directly.

\subsection{Top}

The top quark was discovered in 1995, but now ``Top quarks are everywhere''.~\cite{Kim}
For instance, the LHCb experiment, designed for B physics, recently measured
{\it forward} $t\bar t$ production at 13 TeV.\cite{tt-LHCb}
Or the first observation of top production {\it in p-Pb collisions}\cite{top-pPb CMS} 
by CMS at $\sqrt{s_{NN}} = 8.16$ TeV.
Top quarks are indeed everywhere at the LHC.
Like $M_W$, top mass is a key parameter in SM.
Being the heaviest, it arises from the top Yukawa coupling $\lambda_t \cong 1$,
which affects vacuum stability of our Universe, as well as
sourcing all kinds of loop effects such as in B decays
and aggravating the hierarchy problem (hence motivating ``top partners'').

As for top for its own sake, we touch two {\it top}ics.

\begin{figure}[b]
\vskip-0.5cm\centerline{
\includegraphics[width=5.6cm]{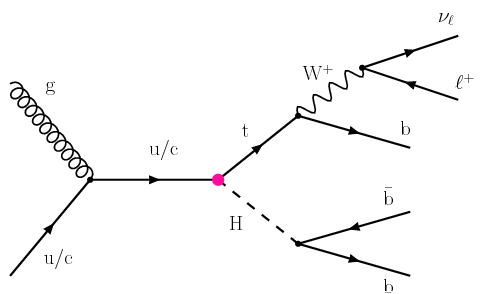}
\includegraphics[width=5.6cm]{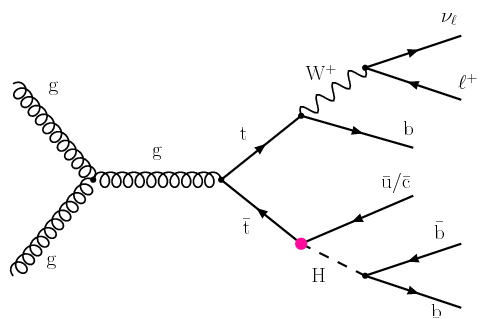}
}
\vskip0.3cm
\caption[]{
Diagrams for singly produced top from $tqH$ coupling,
or $t\bar t$ pair production followed by $t\to qH$ decay,
both with $H \to b\bar b$ [figure taken from Ref.~\citen{tcHbb CMS}].
}
\label{fig:tqHbb}
\end{figure}

One {\it top}ic is spin correlations in $t\bar t$ production and decay.
Since the top quark decays before it hadronizes (there are no ``top mesons''), 
the spin information is preserved.
QCD-produced $t\bar t$ would be unpolarized, but NP could change this.
Thus, spin-correlations, which can be extracted via the leptonic decays of the top pair
($pp \to e\mu b\bar bX$ signature), is in fact a probe of potential NP,
and has been studied since Tevatron days.
A recent result from ATLAS,~\cite{tt-spin-corr-ATL-CONF} 
based on $\sim 36$ fb$^{-1}$ data at 13 TeV,
indicates a rise in differential cross section w.r.t. the angular difference $\Delta\phi$
between the two charged leptons from $W^+W^-$ decay,
with spin correlation larger than SM prediction by 3.7$\sigma$
(3.2$\sigma$ when theory uncertainty is included).
This continues an earlier trend,\cite{Aad:2014mfk,Khachatryan:2016xws} but with better precision.
Let's see how this evolves further, first with the corresponding CMS result,
and with the full Run 2 updates.

\begin{figure}[t]
\centerline{\includegraphics[width=8.2cm]{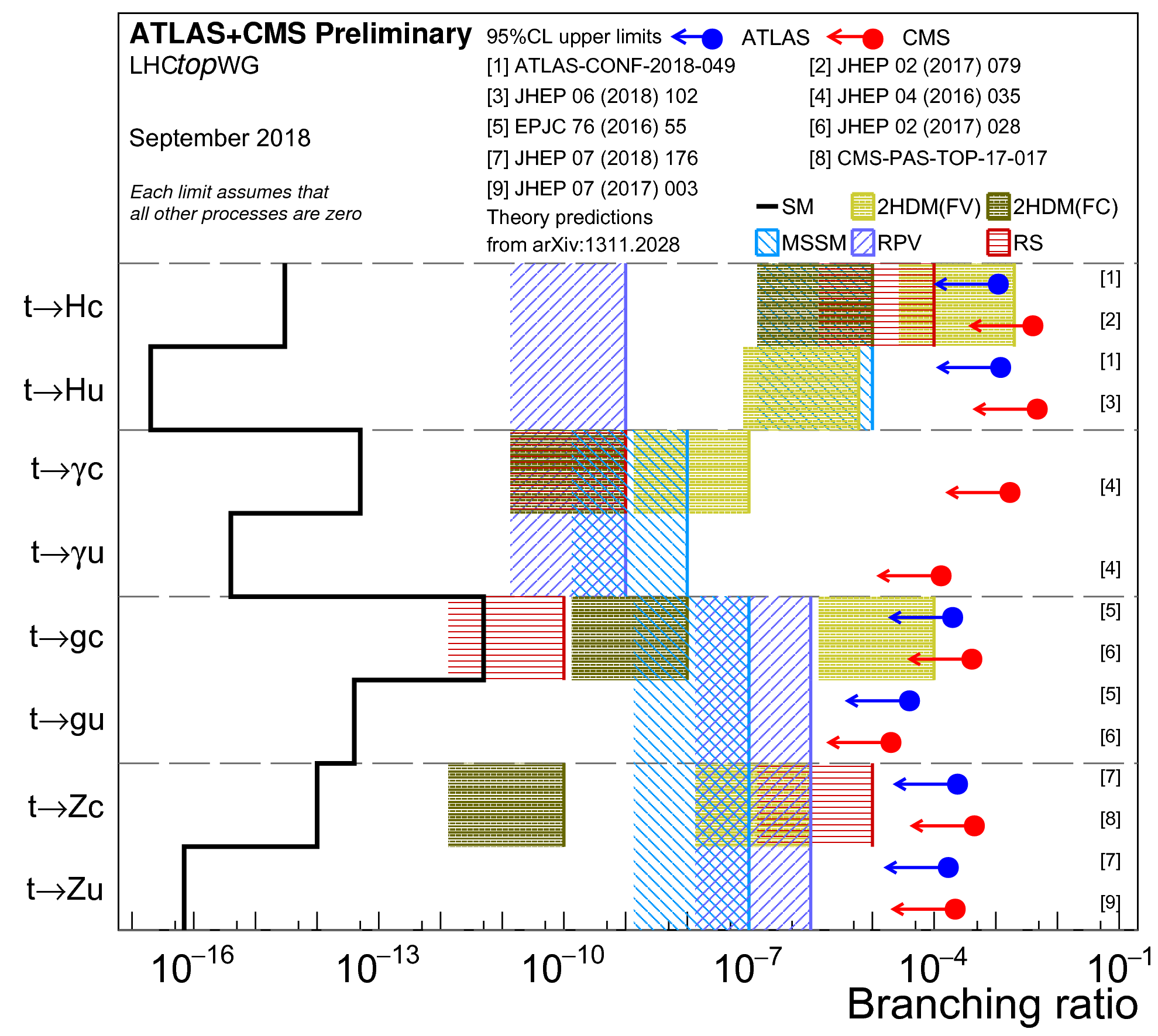}}
\caption[]{
Summary of current search limits for tCNC (top-changing neutral coupling) transitions
[Source: https://twiki.cern.ch/twiki/bin/view/LHCPhysics/LHCTopWGSummaryPlots].
}
\label{fig:tCNC}
\end{figure}

A second, larger {\it top}ic is the search\cite{Kim} for FCNC (Flavor Changing Neutral Couplings).
After the 125 GeV boson was discovered, 
the pursuit of $t\to cH$ has been a highlight interest.\footnote{
It would be better to use the notation $h$ in context when 
more Higgs bosons are implied, but here we stick to notation used by the experiments.}
It is hard to separate a $c$-jet from a light quark jet in actual search.
Further, it is common now to search for $t \to qH$ via both
single top $qg \to tH$ production, as well as usual $gg \to t\bar t$ production
followed by one top decaying via $t \to qH$, where $q$ stands for $u$ and $c$
(see Fig.~\ref{fig:tqHbb}).
This is the case for a recent search by CMS,~\cite{tcHbb CMS} 
using $H \to b\bar b$ decay and based on $\sim 36$ fb$^{-1}$ at 13 TeV. 
The signature is single lepton + 2/3(/4) b jets for $tuH$ ($tcH$) coupling.
Other final states are $h \to \gamma\gamma$,
and multi-leptons (combining $WW^*$, $\tau^+\tau^-$ and $ZZ^*$).
The current best limit, ${\cal B}(t \to cH) < 0.16\%$ at 95\% C.L., 
is by ATLAS~\cite{tcH-mult-l ATL}  
using $\sim 36$ fb$^{-1}$ at 13~TeV with $H \to$ multi-lepton final states.
ATLAS has combined their results for TOP2018,
setting 95\% C.L. bounds for ${\cal B}(t \to cH)$ and ${\cal B}(t \to uH)$
at $1.1 \times 10^{-3}$ and $1.2\times 10^{-3}$, respectively,
with expectation at $8.3 \times 10^{-4}$ for both modes.\cite{ATLAS:2018bka}

The usual FCNC $t\to cZ$ decay has been searched for since top discovery.
With SM expectation far below $10^{-10}$, any discovery would indicate NP.
Although CMS led the way initially,\cite{Chatrchyan:2013nwa} 
the current best limits are from ATLAS:\cite{Aaboud:2018nyl}
${\cal B}(t \to cZ) < 2.4 \times 10^{-4}$ ($3.2 \times 10^{-4}$) and 
${\cal B}(t \to uZ) < 1.7 \times 10^{-4}$ ($2.4 \times 10^{-4}$)
at 95\% C.L., with expectation in parenthesis.
The ultra-rare decays $t \to c\gamma$, $cg$ have also been searched for.

Current tCNC search limits are summarized\cite{Kim} in Fig.~\ref{fig:tCNC}
and compared with theory expectations taken from 2013
Snowmass summer study.\cite{Agashe:2013hma}.
The search for $t\to cH$ is a current frontier where discovery could occur at any time
(more discussion in Outlook).
In contrast, $t \to cZ$ seems to lack NP motivation nowadays.
For example, the extra-dimension or RS model projection (see Fig.~\ref{fig:tCNC}) 
is receding as direct searches place more stringent bounds.

\subsection{Flavor \& CPV}

Flavor physics and $CP$ violation (CPV) is a highlight subject.
We keep this subsection short as we defer the discussion of 
flavor anomalies\cite{Passaleva,Eklund} and the related theory, 
as well as the up and coming Belle II and kaon experiments, 
to the Outlook part.

Here, we just cover very briefly some CPV topics,~\cite{Marks}
where LHCb is pushing the current frontier.
First, the measurement of the CKM phase angle $\gamma$ (called $\phi_3$ by Belle), 
which is the CPV phase of $V_{ub}^*$ in the standard PDG convention,
is now dominated by LHCb, with error $\sim 5^\circ$ at present,
and would continue to improve.
Note that this parameter is as fundamental as the fine structure constant, $\alpha$.
Furthermore, the measurement, based on interference of {\it tree-level} 
$B^+ \to \bar D^{0(*)}K^+$ and $D^{0(*)}K^+$ 
decays to common final states,\cite{Chang:2017wpl}
is a key probe of CKM unitarity that can,
in the limit of large statistics, become free from hadronic uncertainties.

Second, LHCb has found first evidence~\cite{Lam-b-CPV LHCb} 
for CPV in the baryon sector at 3.3$\sigma$ in 
localized asymmetry measurements,
i.e. in the $a_{CP}^{T-{\rm odd}}$ variable measured 
in phase space and $\Phi$-dependent
 (where $\Phi$ is the angle between the two decay planes formed by
  $p\pi^-_{\rm fast}$ and  $\pi^-_{\rm slow}\pi^+$) 
binning of  four-body $\Lambda_b \to p\pi^-\pi^+\pi^-$ decays.
The result is based on 3 fb$^{-1}$ data from Run 1.

Third, LHCb has found {\it no} evidence,~\cite{A_Gam-D0 LHCb} 
down to $10^{-3}$ precision, for indirect CPV in the variables 
$A_\Gamma$ via $t$-dependent study of $D^0 \to \pi^+\pi^-$, $K^+K^-$ decay, namely
$A_\Gamma(K^+K^-) = (-0.30 \pm 0.32 \pm 0.10) \times 10^{-3}$ and
$A_\Gamma(\pi^+\pi^-) = (0.46 \pm 0.58 \pm 0.12) \times 10^{-3}$.
Thus, there is no evidence so far for $CP$ violation in the charm sector,
which is in itself not too surprising.
But one should recall the backdrop, that LHCb had once found evidence
in early data (0.62 fb$^{-1}$) for {\it direct} CPV difference, $\Delta A_{\rm CP}$,
between the $D^0 \to \pi^+\pi^-$, $K^+K^-$ decay modes 
at the \% level,\cite{Aaij:2011in} which caused some sensation,
but unfortunately turned out to be a fluctuation.\cite{Aaij:2014gsa}

\subsection{Spectroscopy: XYZ States and Others}

Charmonium-like XYZ particles~\cite{Shen} started with 
the Belle discovery\cite{Choi:2003ue} of X(3872) in 2003,\footnote{
The observation of X(3872) is in fact the best cited Belle paper.
}
opening the window to multiquark hadrons,
which has flourished as a subfield since.
Sample reviews are Refs.~\citen{XYZ17} and \citen{Karliner:2017qhf}.
At the {\it Rencontres du Vietnam}, BESIII reported for the first time~\cite{Shen} 
the 5.2$\sigma$ observation of X(3872)$\to \pi^0 \chi_{\rm c1}$ decay mode,
with no evidence involving $ \chi_{\rm c0}$ or $\chi_{\rm c2}$ in final state,
which disfavors the $ \chi_{\rm c1}(2P)$ interpretation of X(3872).

What makes the case for four-quark (or molecular) states
even more compelling are the ``charged charmonium'' states,
such as Z$_{\rm c}$(3900) observed by Belle and BESIII in 2013.
For more detailed discussion, see Ref.~\citen{XYZ17}.
At the {\it Rencontres du Vietnam}, BESIII also reported for the first time~\cite{Shen} 
strong evidence, at 4.3$\sigma$ statistical significance, 
the Z$_{\rm c} \to \rho^-\eta_{\rm c}$ decay mode
in $e^+e^- \to \pi^+\pi^-\pi^0\eta_{\rm c}$ production.
In a recent paper published by D0,~\cite{Zc-D0}
strong evidence (at 4.6$\sigma$) was found~\cite{Hirosky,Shen} 
for Z$_{\rm c}$ production in $b$-hadron decays to $Y(4260)$, 
the famed $1^{--}$ state discovered by BaBar.\cite{Aubert:2005rm}
That is, Y(4620)$ \to$ Z$_{\rm c}^\pm \pi^\mp$ with 
Z$_{\rm c}^\pm \to \pi^\pm J/\psi$.

On a somewhat negative note, the X(5568) state that was 
claimed~\cite{5568-D0} by D0 may not be there.
D0 claimed strong evidence (4.8$\sigma$) for X(5568) $\to B_s^0\pi^\pm$ decay
in $B_s^0 \to J/\psi \phi$. If true, this would be the first
tetraquark with all four different flavors.
However, besides further support~\cite{Hirosky} from D0 in
 $B_s^0 \to \mu D_s +X$ semileptonic decay,
searches~\cite{Shen} by LHCb,\cite{Aaij:2016iev} 
CMS,\cite{Sirunyan:2017ofq} CDF\cite{Aaltonen:2017voc} 
and ATLAS\cite{Aaboud:2018hgx} all turned out negative.

\begin{figure}[b]
\hskip0.2cm
\begin{minipage}{0.45\linewidth}
\centerline{\includegraphics[width=0.9\linewidth]{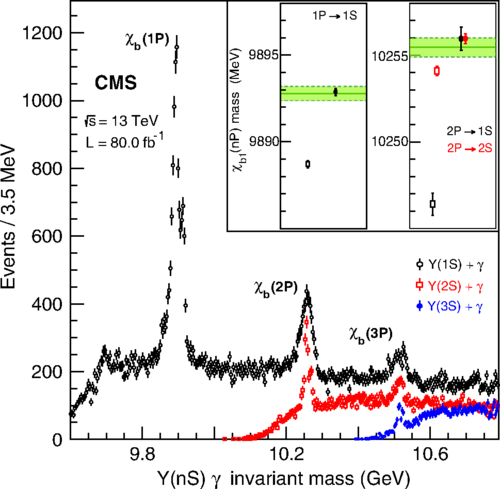}}
\end{minipage}
\hskip0.05cm
\begin{minipage}{0.5\linewidth}
\centerline{\includegraphics[width=0.9\linewidth]{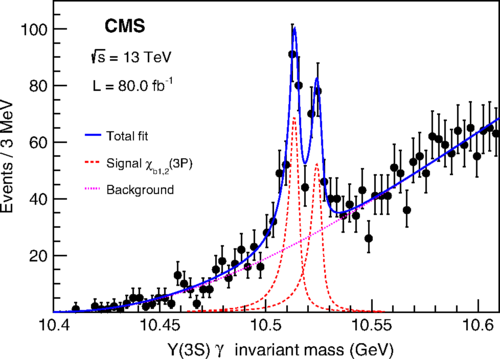}}
\end{minipage}
\caption[]{
 Mass distribution for [left] $\chi_{bJ} \to \Upsilon(nS)\gamma$, and
 [right] $\chi_{bJ}(3P) \to \Upsilon(3S)\gamma$.\cite{Sirunyan:2018dff}
}
\label{fig:dm-chibJ-3}
\end{figure}

To show the vitality of the field of heavy flavor spectroscopy and the prowess of LHC experiments,
we note some recent results by LHCb\cite{Passaleva} and CMS:\cite{Butler}
\begin{itemlist}
 \item Doubly charmed baryon $\Xi_{cc}^{++}$:
   Discovered only in 2017 by LHCb via $\Xi_{cc}^{++} \to \Lambda_c^+ K^-\pi^+\pi^+$,
   the experiment has measured the lifetime\cite{Aaij:2018wzf} 
   and uncovered\cite{Aaij:2018gfl} a second channel, $\Xi_{cc}^{++} \to \Xi_c^+\pi^+$.
   All these results are based on 1.7 fb$^{-1}$ from Run~2.
 \item $\Omega_c^0$ lifetime: 
   One of the striking results for summer 2018 is the LHCb measurement,
   based on 3 fb$^{-1}$ from Run 1,
   finding $\tau_{\Omega_c}|^{\rm LHCb} \simeq 268$ fs,\cite{Aaij:2018dso}
   which is {\it four times} the PDG value of $\tau_{\Omega_c}|^{\rm PDG} \sim 69$ fs!
   This definitely needs to be digested by theory.
 \item New $\Xi_b(6227)^{-}$ resonance: 
   Observed\cite{Aaij:2018yqz} by LHCb in $\Xi_b^0\pi^-$ and
   $\Lambda_b^0K^-$ final states,
   with 3 fb$^{-1}$ from Run 1 plus 1.5 fb$^{-1}$ from Run 2.
 \item Resolved $\chi_{b2}(3P)$--$\chi_{b1}(3P)$ mass splitting:\cite{Sirunyan:2018dff}
   The  $\chi_{b}(3P)$ state, first observed around 10.5 GeV by ATLAS,\cite{Aad:2011ih}
   is closest to the $b\bar b$ continuum, and could be\cite{Karliner:2014lta}
   the ``$X_b$'' state\cite{Hou:2006it} that corresponds to $X(3872)$.
   With its strong 3.8T magnetic field 
   and using 80 fb$^{-1}$ of Run 2 data,
   CMS was able to resolve the $J=1$ and 2 states
   via $\chi_{bJ}(3P) \to \Upsilon(3S)\gamma$, with $\Upsilon(3S) \to \mu^+\mu^-$
   and $\gamma \to e^+e^-$ (conversion in silicon tracker).
   Individual masses are measured at 
   $m_{\chi_{b1}(3P)} = 10513.42 \pm 0.41 \pm 0.18$ MeV and
   $m_{\chi_{b2}(3P)} = 10524.02 \pm 0.57 \pm 0.18$ MeV, with 
   mass splitting at $10.60 \pm 0.64 \pm 0.17$ MeV
   (see Fig.~\ref{fig:dm-chibJ-3}).
\end{itemlist}

\subsection{Dark Matter}

DM is a vast subject at the intersection of astrophysics and particle physics
that begs for improved understanding. 
We can only briefly touch the particle physics side,
and what we present should be viewed as ``windows'' on subjects of pursuit.

Let us start with the classic WIMP (Weakly Interacting Massive Particle) search 
at the LHC, then direct search with novel approaches, and then into alternatives.

\begin{figure}
\centerline{\includegraphics[width=5cm]{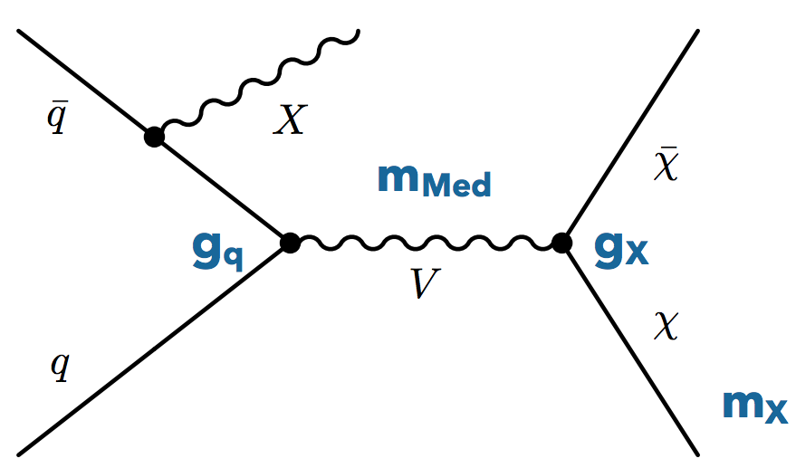}}
\caption[]{
 Simplified model for mono-X search approach for DM at LHC.~\cite{Kouskoura,DM-Run2}
}
\label{fig:DM-LHC}
\end{figure}

\subsubsection{WIMP: DM @ LHC\cite{Kouskoura}}

DM search at LHC follows the WIMP paradigm, 
largely utilizing missing energy and mass, or ``mono--X'' processes,
where $X$ is some tag particle, such as
$q/g$-jet,\cite{Aaboud:2017phn,Sirunyan:2017jix} photon, $W/Z$, 
$b/b\bar b$ or $t/\bar t$,\cite{Aaboud:2017rzf,Sirunyan:2018dub} 
Higgs.\cite{ATLAS:2018bvd,Sirunyan:2018fpy,Sirunyan:2018gdw}
A simplified model approach is adopted,~\cite{DM-Run2} 
parametrized by 4 free parameters (see Fig.~\ref{fig:DM-LHC}):
mediator and DM masses $m_{\rm Med}$ and $m_\chi$, 
and mediator couplings to quarks ($g_q$) and DM ($g_{\chi}$).

A wide range of searches are pursued by ATLAS, CMS and LHCb
with no excess observed so far, placing bounds on DM and mediator masses,
including axial-vector and (pseudo-)scalar mediators.
Some of the bounds can be found in Figs.~\ref{SUSY18} and \ref{Exotics18}.
New developments, such as emerging jets~\cite{Butler} 
based on LLP type of ideas, are also being explored.
An example\cite{Sirunyan:2018njd} is new particle search via ``a jet and an emerging jet'',
where the emerging jet corresponds to multiple displaced vertices,
i.e. multiple tracks with large impact parameter.
We have also discussed in Sec. 1.4 the related long-lived 
(hence displaced vertex) Dark Photon search by LHCb,\cite{Dphot-LHCb}
and mentioned there in a footnote a recent search for low mass DM by Belle,\cite{Seong:2018gut}
to illustrate the broad interest.
 
Our discussion is rather incomplete, and the analyses are evolving
towards full Run 2 data. But the main point is that, unfortunately, 
there are no hints so far for any DM candidates from LHC searches,\cite{Kouskoura}
or other accelerator searches.

\begin{figure}
\centerline{\includegraphics[width=8.5cm]{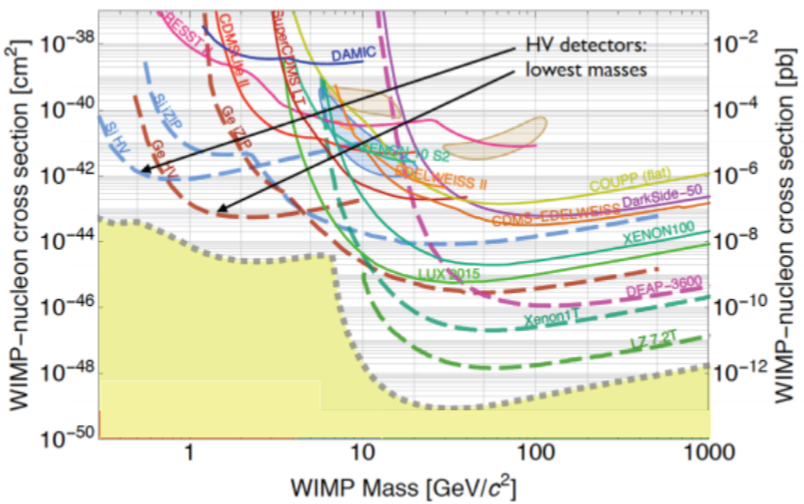}}
\caption[]{
 Current and projected direct search bounds [Source: P. Cushman (2017)].~\cite{Lopes}
}
\label{fig:DM-Direct}
\end{figure}

\subsubsection{WIMP: Direct Detection\cite{Lopes}}

As evidence for DM is gravitational, we live actually in the galactic DM cloud.
Direct detection of WIMP particles is based on the recoil of some nucleus
from collision with cosmic DM particle, $\chi$.
Because of our lack of understanding of the true nature of the DM particle(s),
including spin, there is room for ingenuity, and DM direct detection has 
flourished into a global enterprise with many experiments,\cite{DM-Run2} 
including in East Asia and Australia (SABRE).
The current bounds are illustrated by solid lines in Fig.~\ref{fig:DM-Direct},
with dashed lines projecting into the future.
This is quite exquisite a field, which we cannot go into any detail.

Two boundaries are being pushed. 
One is the Xenon-based large ($> 1$ ton) detectors,
such as LUX/LZ,\cite{Akerib:2018lyp}
 XENON,\cite{Aprile:2018dbl}
 PandaX,\cite{Ren:2018gyx} 
aiming for large exposure time.
These would push down in traditional WIMP mass range, 
towards the atmospheric neutrino background ``floor''.
But the relatively heavy Xenon loses sensitivity for DM mass 
below $\sim$ 10 GeV.\footnote{ 
 Hence recent developments in liquid Argon, such as DarkSide,\cite{Agnes:2018fwg}
 which explores far less radioactive underground Argon sources,
 aiming eventually for background-free kton-year searches.
}
This less explored region is the realm of cryogenic semiconductor 
(such as Silicon or Germanium) detectors pushing for 
low thresholds and amplified signals, 
such as SuperCDMS\footnote{
 SuperCDMS\cite{Agnese:2018col} aims for moving to
 SNOLAB (https://www.snolab.ca/) with iZIPs 
 (interleaved Z-sensitive Ionizaation and Phonon) Ge detectors,
 and to start operation ca. 2020.
} 
or CRESST.\cite{Petricca:2017zdp}
We also see the development of ultra-pure Ge detectors,
such as CDEX\cite{Jiang:2018pic} at the deep 
China Jinping Underground Laboratory, where PandaX is also located.

Business for DM is by far not yet finished, but looking at 
the coverage in Fig.~\ref{fig:DM-Direct},
the absence of a signal bears resemblance to the situation at the LHC.
Are we on the right track?

\subsubsection{FIMP alternative,\cite{Tytgat}  Search for milli-Q,~\cite{Hill}
   and Other Approaches }

Absence of any signal so far for WIMPs, both at the LHC and in direct detection,
has stimulated thoughts for non-WIMP scenarios, such as ``FIMP''.\cite{Tytgat}

The FIMP idea is based on a dark photon $\gamma^\prime$
(often denoted as $A'$, as is the case for\cite{Dphot-LHCb,Ilten:2016tkc} 
the LHCb inclusive $A' \to \mu^+\mu^-$ search)
that provides a portal to some hidden sector, where small kinetic mixing\cite{Holdom:1985ag}
with the photon gives rise to some ``hidden'' {\it milli-charged} particles of unknown mass. 
Thus, instead of the WIMP freeze-out,
cosmic abundance of DM is built up from slow production,
which is called\cite{Tytgat} Freeze In (FI). 
In this case, traditional direct detection
tests the idea in $t$-channel, and could be enhanced
for light $\gamma^\prime$.~\cite{Hambye:2018dpi}

If DM are milli-charged, hidden sector particles, they act as 
MIPs with very feeble $dE/dx$ and can traverse long distances.
Proposed experiments such as milliQan\cite{Hill,Haas:2014dda}  can detect
such particles produced in the CMS detector, 
passing through bedrock and well shielded from cosmic rays.
A 1\% demonstrator was installed in 2017, and has been running since.\cite{Hill} 
If funding arrives in time, a full-scale experiment could be 
installed during LS2, and run for Run 3 and beyond.
We mention also the Light Dark Matter eXperiment (LDMX) 
concept,\cite{Akesson:2018vlm}
which covers other Light DM particles and mediators as well.
For neutral Long-Lived Particles, we have already mentioned
MATHUSLA\cite{Alpigiani:2018fgd} in Sec.~1.4.
There is also MoEDAL at the LHCb site (Point 8),
searching\cite{Acharya:2017cio} for monopoles and other exotics.

Our coverage here is of course far from complete,
and DM search is certainly not limited to WIMPs or even FIMPs.
A very popular field of research is axion DM and its extension
to ALPs (Axion-Like Particles). 
We refer to the short review of Ref.~\citen{Irastorza:2018dyq}
for the diverse and very active experimental approaches.
Not to be forgotten are the {\it indirect} detection searches
of astro-DM annihilation, such as the famous positron excess
observed by PAMELA\cite{Adriani:2008zr} and 
followed up by AMS.\cite{Aguilar:2013qda}

Understanding the nature of DM is of utmost importance, 
and LHC search is only part of a very broad program,
which goes beyond traditional HEP.

\subsection{Neutrino}

We group present and future long baseline neutrino experiments, 
short baseline neutrino experiments, as well as 
neutrinoless double beta decay ($0\nu\beta\beta$)
under the banner of neutrino physics.

\subsubsection{NOvA\cite{Choudhary}}

NOvA is a second-generation long baseline ($\sim 800$ km) 
experiment on the NuMI beamline from Fermilab, 
and is optimized for the detection of $\nu_\mu \to \nu_e$ oscillations.
It took neutrino data with $8.85 \times 10^{20}$ POT (protons on target) up to early 2017,
the result of which was published recently.~\cite{NOvA:2018gge}
Anti-neutrino data taking continued until summer 2018, 
but a preliminary joint analysis of neutrino and anti-neutrino data 
(corresponding to $6.9 \times 10^{20}$ POT) was reported
at {\it Rencontres du Vietnam}.\cite{Choudhary}
NOvA finds $> 4\sigma$ evidence for $\nu_e$ appearance,
the first such result for this channel.
Less significantly, NOvA prefers normal mass hierarchy (NH)
at 1.8$\sigma$ level, and excludes $\delta_{CP} = \pi/2$ 
for the $CP$ phase at 3$\sigma$ for inverted hierarchy (IH).
Future running can reach 3$\sigma$ sensitivity 
for normal mass hierarchy by 2020 if one has $\delta_{CP} = 3\pi/2$, 
and will cover a significant range of $\delta_{CP}$ by 2024.

\begin{figure}
\centerline{\includegraphics[width=5.5cm]{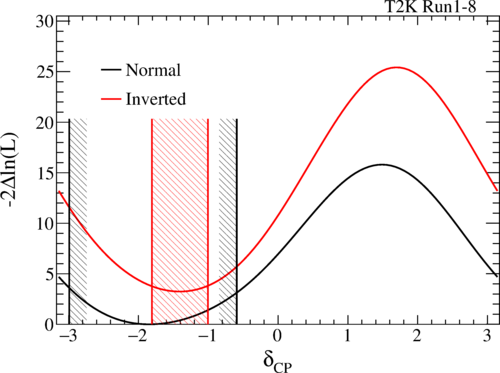}}
\caption[]{
 Log likelihood for $\delta_{\rm CP}$ by T2K,\cite{Abe:2018wpn}
 with best fit at near maximal CPV phase.
}
\label{fig:delCP-T2K}
\end{figure}

\subsubsection{Future Long Baseline Experiments}

More intense beam power and larger detectors are needed for the next
generation long baseline experiments to measure CPV,
where effects are more pronounced at lower energies.
Two main experiments are planned:
\begin{itemize}
 \item HK: baseline of 295 km from J-PARC 
    using water Cherenkov technology;\cite{DiLodovico}
 \item DUNE: baseline of 1300 km from Fermilab, using liquid Argon technology.\cite{Ereditato}
\end{itemize}
Based on the success of Kamiokande and SuperK, which led to two Nobel prizes,
it is natural to pursue an ever larger detector.
Tokyo University announced~\cite{HK-todai} in September 2018 for receiving 
``seed money'' for HK (Hyper-Kamiokande), which by Japanese tradition implies that 
funding could start in 2019. Construction could then start in 2020, 
with data taking slated for 2026.~\cite{DiLodovico}
This follows the proposal\cite{Abe:2016tii} to extend T2K running
 (T2K-II) to $20 \times 10^{21}$ POT, 
with continuous beam power increase to 1.3~MW for HK.
The physics aims cover~\cite{DiLodovico,Abe:2018uyc} 
exclusion of $\delta_{CP} = 0$: with a decade of running,
 80\% coverage of $> 3\sigma$ exclusion, and can reach
 $8\sigma$ for $\delta_{CP} = -\pi/2$, the current T2K best fit\cite{Abe:2018wpn}
 for Normal Hierarchy (see Fig.~\ref{fig:delCP-T2K}, lower curve).
By combining atmospheric and beam data, HK can determine mass hierarchy,
and it will of course also address proton decay anew.\footnote{
 It certainly would be nice to place a second HK in Korea,\cite{Abe:2016ero} 
 downstream from HK in Japan, i.e. to have T2/HKK.
 But how this would materialize remains to be seen.
}

DUNE (Deep Undergound Neutrino Experiment) and LBNF (Long Baseline Neutrino Facility)
is the Fermilab answer\cite{Ereditato,Acciarri:2015uup} to HK and T2HK, 
which arose after long soul-searching on how to transform itself 
as the only remaining dedicated HEP lab in the US.
Since the DUNE Conceptual Design Report (CDR), 
the Near Detector (ND) design is now approaching final,
with a CDR targeted for 2019.
The Far Detectors would consist of $4\times 10$ kton LAr
Time Projection Chambers (TPC), where 
two 1/20 scale ``proto-DUNE'' detectors (single and dual phase) 
have been constructed and tested recently at CERN
 (see {\it CERN Courier} October 2018 issue).
DUNE TDR is expected in 2019, with schedule and target similar to HK.
The beam power of 1.2 MW is upgradable to 2.4 MW.

As a (heavy) flavor physics person, let me state, in envy: 
In contrast to the hierarchical pattern reflected in the CKM matrix, 
neutrino physics is privileged with three large mixing angles
in the PMNS matrix, making $\delta_{CP}$ relatively accessible.
But the interconnection between quark and neutrino flavor
is not understood.
Are the CKM and PMNS matrices related in any way? 
What about charged fermion vs neutrino masses?

CPV measurement, however, is further down in the timeline.
The current race is for the aforementioned neutrino mass hierarchy:
we know that $\nu_1$--$\nu_2$ are closer to degeneracy,
but is $\nu_3$ higher (NH) or lower (IH)?
Let us mention another worthy contender in this race:\cite{An:2015jdp} 
JUNO (Jiangmen Underground Neutrino Observatory).
JUNO, a 20 kton liquid scintillator detector located in
the Guandong province of China, is the successor to
the successful DayaBay Neutrino Experiment
which led the discovery,\cite{An:2012eh} also in 2012, 
of sizable $\sin^2\theta_{13} \simeq 0.09$,
the third neutrino mixing angle.
Benefiting from this, JUNO is fully funded and under construction,
with start of data taking aimed for 2020.

\subsubsection{Short Baseline Experiments\cite{Karagiorgi}}

Interest here relates to the question of the possible existence of sterile neutrinos,
beyond the three known active ones, which does matter for the Universe.
Experimentally, it traces back to the LSND experiment in the 1990's, 
then MiniBooNE and reactor neutrino measurements,
giving a chain of anomalous excesses of $\nu_e$ in $\nu_\mu$ beam, 
or $\nu_e$ deficits from $\nu_e$ sources, typically at $L/E \sim$ 1 m/MeV.

\begin{figure}[b]
\centerline{\includegraphics[width=7.5cm]{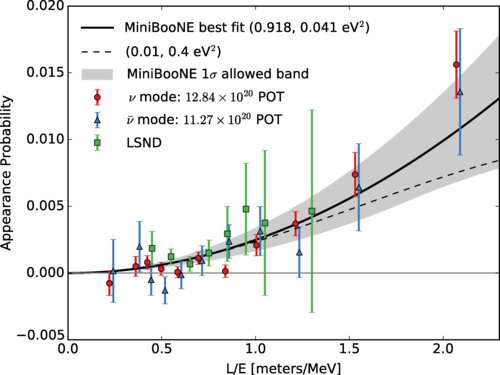}}
\caption[]{
Consistency between MiniBooNE $\nu$ and $\bar \nu$ data
with LSND.\cite{Aguilar-Arevalo:2018gpe} 
}
\label{fig:mBooNE18}
\end{figure}

The ongoing MiniBooNe experiment announced recently\cite{Aguilar-Arevalo:2018gpe} 
an excess of $\nu_e$-like events at too short a distance,
and is consistent with a sterile neutrino interpretation of the old LSND result
(see Fig.~\ref{fig:mBooNE18}), where the two experiments have quite different systematics.
But, of course, not everything fits perfectly on this subject.\cite{Kopp:2018rbc}
More sensitive tests~\cite{Karagiorgi} are coming soon from reactor-based
 SoLiD,\cite{Abreu:2018pxg}
 DANSS,\cite{Alekseev:2018efk}
 NEOS,\cite{Ko:2016owz}
 STEREO,\cite{Almazan:2018wln}
 PROSPECT\cite{Ashenfelter:2018iov} experiments,
and at the Fermilab Short Baseline Neutrino (SBN) program,\cite{SBN}
where a trio\cite{Antonello:2015lea,Cianci:2017okw}
 (SBND, MicroBooNE\cite{Acciarri:2016smi} and ICARUS) of 
accelerator-based experiments using LAr TPCs
are coming online by 2019--2020.
The large number of experiments illustrates the keen interest.

The sterile neutrino landscape will certainly be scrutinized further!

\subsubsection{$0\nu\beta\beta$: Quest for Majorana Neutrinos\cite{Schoenert}}

In terms of experimental methodology, the pursuit of neutrinoless double beta decay ($0\nu\beta\beta$)
measurement overlaps with some of the approaches with DM direct search.

Neutrinoless double beta decay can occur if the neutrinos are
Majorana particles, i.e. their own antiparticles.
The GERDA experiment reported recently~\cite{Agostini:2018tnm} 
new results on background-free search for $0\nu\beta\beta$, reaching
half-life sensitivity at $8 \times 10^{25}$ ($\sim 10^{26}$) yr at 90 \% C.L.
Together with Majorana Demonstrator, linear improvement is expected,
and as reported at the {\it Rencontres du Vietnam},\cite{Schoenert}
the next generation experiments (e.g. LEGEND\cite{LEGEND}) 
aim at a 100 fold increase in sensitivity, reaching $10^{28}$ yr!

This definitely should be watched and followed.\footnote{
 We note that the ambitious tritium beta decay experiment,
 KATRIN, has started\cite{newsKATRIN} data taking in 2018,
 which would continue for 5 years.
 The aim is to measure $m_{\bar\nu_e}$ with sub-eV sensitivity.\cite{Angrik:2005ep}
}

\subsection{H.E. Universe: from IceCube to Theory}

We group the recent IceCube result and a few theory topics
under the {\it H.E. Universe}.

\subsubsection{IceCube and the H.E. Universe\cite{Resconi}}

IceCube is an instrumented km$^3$ cube of ice at the South Pole,
a multipurpose detector with main aim for cosmic neutrino detection.
A pair of {\it Science} articles highlight some recent observations.
A high-energy ($\sim 290$ TeV) muon neutrino track event, IceCube-170922A, 
points back~\cite{IceCube:2018cha} to a known $\gamma$-ray blazar, TXS 0506+056.
With direction and time window known, this triggered a multi-messenger
confirmation~\cite{IceCube:2018dnn} of the blazar being in a flaring state,
and IceCube finding a dozen of H.E. neutrinos (3.5$\sigma$) 
during 2014--2015.~\cite{IceCube:2018cha} 
Thus, blazars appear to be a source of astrophysical $\nu$'s.
Though not HEP {\it per se}, it offers a startling {\it Window on the H.E. Universe!}

IceCube itself has contributed to cosmic and atmospheric neutrino studies, 
astro-DM search, and a host of other interesting HEP topics.

\subsubsection{Theory and the H.E. Universe}

As a snapshot, we mention briefly some theory topics 
covered at the {\it Rencontres du Vietnam}
(again, flavor anomaly discussion is deferred to Outlook):
\begin{itemize}
\item Pushing the {\it Emergence} Frontier:\footnote{
 During {\it Rencontres du Vietnam},
 ICTP announced that Prof. Dam Thanh Son,
 the speaker on this subject, received the Dirac Medal,
 for work that crosses high energy and condensed matter
 (even nuclear) physics boundaries.
 Congratulations, and even better works to come!
 }
   Using half-filled Landau levels in condensed matter physics
   to illustrate\cite{Son} the {\it Emergence} of composite Dirac fermions,
   and implications for HEP, duality, etc.\cite{Son:2018cba} 

\item Inflation as Cosmological Collider:\cite{Wang}

   Viewing inflation as a cosmological collider at the Hubble scale,
   characteristics of very massive particles could be recorded
   in the primordial non-Gaussianities.\cite{Chen:2016uwp}

\item Gravitational Waves (GW) as Probe of Early Universe:\cite{Figueroa}

   GWs from first order phase transitions (collision of bubble walls) and cosmic defects
   are detectable by detectors such as LISA,
   which can probe both the Early Universe, as well as High Energy Physics.
   See Ref.~\citen{Caprini:2018mtu} for a review.

\item Ideas on Origin of the Weak Scale:\cite{Dudas}

   Newer ideas of twin Higgs,
   relaxion (making Higgs mass dependent on the axion field), and
   ``Intelligent Ultraviolet Completion'' (IUVC).\cite{Dudas}
   The latter is high-scale SUSY where Higgs mass is
   protected by a symmetry invisible from 4D.\cite{Buchmuller:2018eog}
\end{itemize}

\subsection{ILC/CEPC: Asia!(?)}

There is no doubt that Asia is ascendant economically,
but it may also be ascendant in world HEP, or the HEP world.
This is especially potent an issue at this 25th anniversary of the {\it Rencontres du Vietnam}.
How would things look at the 50th anniversary!

The most important thing,~\cite{Gao} at time of {\it Rencontres du Vietnam}, 
is to hear the positive decision by end of 2018 from the Japanese government on
whether, or not, to host ILC250 in Japan,\footnote{
 At the last iteration of the ILC (International Linear Collider) saga,
 ca. 2016--2017,
 the proponents decided to shorten the length to reduce (initial) cost,
 relying on the extendability, and progress in klystron or acceleration technology,
 for future extensions.
}
and start the international negotiations. 
Signs at the dedicated LCWS2018 conference held October 
in Arlington, Texas, looked promising.
The ideal timeline would be construction start in $\sim 5$ years, 
and completion by another 9 years.
Unfortunately, the Science Council of Japan issued a negative report in Decemeber 2018.
But all is not lost, as the final decision is with the Japanese government,
and things in Japan are ``subtle'', i.e. often not as it seems on the surface. 
MEXT of Japan has announced that the decision would be deferred until before 
the joint meeting of LCB (Linear Collider Board) and ICFA
 (International Committee for Future Accelerators),
to be held early March 2019 in Tokyo.

\begin{figure}[t]
\centerline{\includegraphics[width=0.7\linewidth]{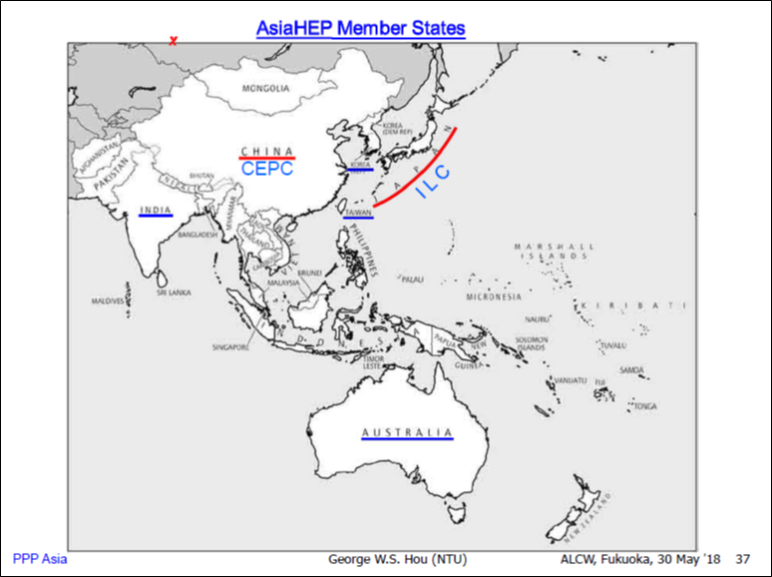}}
\caption[]{
ILC, CEPC, and AsiaHEP member states
(including Novosibirsk, the ``{\scriptsize $\times$}'').}
\label{fig:AsiaHEP}
\end{figure}

The world is clearly waiting and watching for Japan's decision on the ILC.
The other large Asian economy, China, has proposed~\cite{Gao} 
the CEPC-SPPC, which is the 100 km analog of LEP-LHC,
i.e. a Super-LEP/LHC complex, starting with 
the CEPC (Circular Electron Positron Collider) ``Higgs Factory''.
Since the 2015 Pre-CDR, or the preliminary studies, the CDR was finally 
completed in 2018.\cite{CEPCStudyGroup:2018rmc,CEPCStudyGroup:2018ghi}
The next step is the TDR, and ideally construction could start
in 2022, taking also $\sim 9$ years towards completion.

It would be nice if CEPC and ILC could run in parallel initially
as Higgs factories to crosscheck each other.
But they would take different evolution paths.
If {\it both} the ILC and CEPC could be realized, Asia would
definitely take center stage in world HEP by the 50th anniversary of the {\it Rencontres du Vietnam}!
We give in Fig.~\ref{fig:AsiaHEP} the map of the current AsiaHEP member states.\footnote{
 As of 2018:  Australia, China, India, Japan, Korea and Taiwan, plus Novosibirsk.
 }
The amalgamation of ACFA and AsiaHEP is planned for early 2019,
to make it closer to ECFA in structure, composition and mission.
With its economy fast rising, Vietnam is welcome to join when
its HEP community matures.


\section{ Perspective and Outlook }

LHC Run 2 has ended, and Long Shutdown 2 has began.
Judging from the past, the ``Run 2 Era'' extends well beyond LS2 into at least 2021.
With data increase by a factor of 5 from Run~1 and almost double the collision energy to 13 TeV, 
there is much to look forward to.
Besides LHC data-mining, with turn-on and data collection at
many new facilities and experiments, especially along the flavor front, 
{\it the Outlook is Bright (and Flavorful)}.

\subsection{No New Physics: SM Checks Out,Again}

One thing impressed me greatly in summer 2011:\cite{Bachacou}
\begin{quote}
``Unfortunately, no hint of New Physics in the LHC data (yet).''
\end{quote}
This was up to 1/2/3 TeV mass bounds, 
depending on the type of New Physics.
Unfortunately, to date the statement rings true, 
and there is {\it Angst} at the LHC!
Take SUSY for example.
To quote Dan Piraro (Bizarro comics: elephant in the room):
\begin{quote}
``If you were in the middle of the room the whole time,
why can we not find a single witness to corroborate 
your testimony?''
\end{quote}
The same holds true for {\it any} NP.
Note that the bulk of ``the room'' has been quickly scanned already.
So, the search continues, but in Cracks?
This reminds us of the 1)--5) vibrant search areas mentioned in
Sec.~1.3 of our {\it Summary and Highlights} (see Fig.~\ref{SUSY18}).
As NP with SM coupling strength seems exhausted up to a few TeV,
there may still be particles out there with couplings
weaker than in SM (and they need not be LLPs!),
and that is why we have to run the full course of the HL-LHC,
to gain the statistical power.
We must dutifully walk the walk, and we may get rewarded that way.
Who knows.

\vskip0.15cm
So, {\it No New Physics} in sight, and SM  checks out real well.
Along this line, the top three highlights at the LHC so far are (to me):
\begin{itemlist}
\item Discovery of $h(125)$ in 2012 by ATLAS and CMS,\cite{ATL-h,CMS-h}
 and that it so resembles\cite{Khachatryan:2016vau} the SM Higgs boson.

 If SUSY is not seen, ``No Higgs'' is definitely gone, and we have stressed
 the remarkable proximity of $h$ to SM-Higgs, the mysterious
 alignment phenomenon.
 What would full Run 2 data reveal to us?
\item Finding $B_s$-mixing CPV phase $\varphi_s \sim 0$ by LHCb\cite{LHCb:2011aa} in 2011,
 prior to the Higgs boson discovery.
 One should recall that there were some good hope,\cite{Holdom:2009rf} 
 both theoretical and experimental, even at the Tevatron, 
 for a sizable $\varphi_s$ that deviated significantly from SM.

 With the hope dashed, it has been, and would be, slow motion towards
 measurement of $\varphi_s|^{\rm SM}$.
 There still might be discovery along the way.
\item Observation\cite{CMS:2014xfa} of the very rare
 $B_s \to \mu^+\mu^-$ decay in 2015, by combining CMS and LHCb Run 1 data,
 which LHCb could reaffirm as single experiment in 2017
 by adding 1.4 fb$^{-1}$ Run 2 data.\cite{Aaij:2017vad}

 Though the end result may disappoint, it is the crowning glory of the high-stakes saga,
 especially at the Tevatron, that SUSY or the related 2HDM type II could have\cite{Chang:2017wpl}
 greatly enhanced $B_s \to \mu^+\mu^-$ by $\tan^6\beta$.
 We will have to see whether $B_d \to \mu^+\mu^-$ can be
 observed with full Run 2 data.
\end{itemlist}

\begin{figure}[t]
\vskip-0.5cm\centerline{
\includegraphics[width=5.2cm]{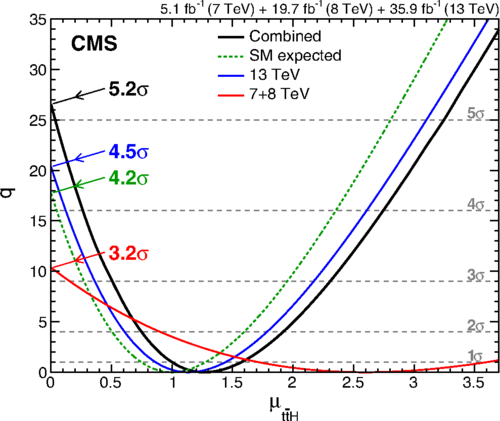}
\includegraphics[width=6.2cm]{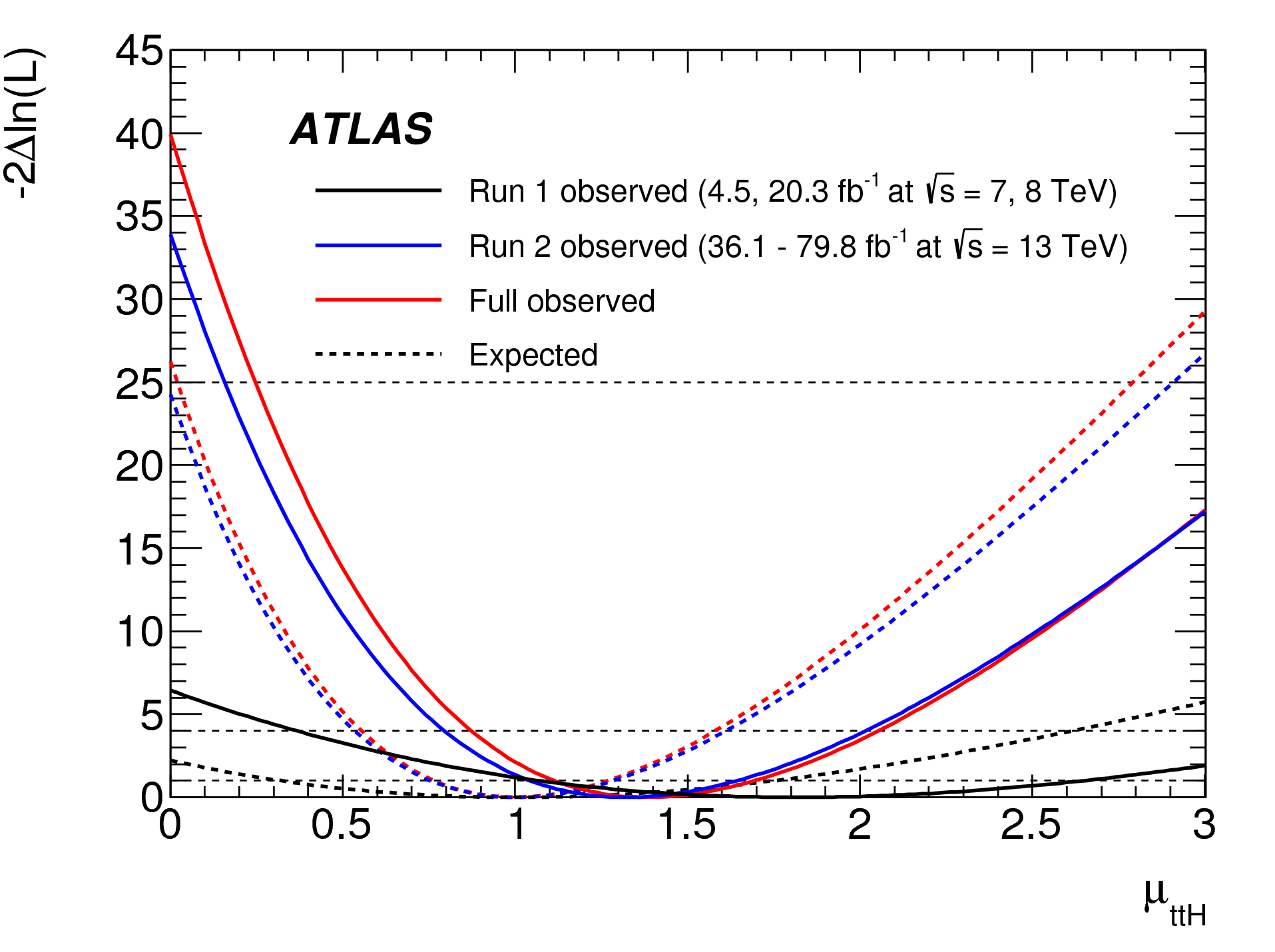}
}
\caption[]{
Observation of $pp \to t\bar tH$ production by
CMS and ATLAS (from Refs.~\citen{Sirunyan:2018hoz} and \citen{Aaboud:2018urx}).
}
\label{fig:tth}
\end{figure}

And the highlight of 2018? To me it is the observation of $t\bar tH$,
by both CMS  and ATLAS (see Fig.~\ref{fig:tth}).
By fate --- help of ``the red line'', the largish Run 1 result at $3.2\sigma$ --- 
CMS could publish\cite{Sirunyan:2018hoz} 
with $\sim 36$ fb$^{-1}$ data at 13 TeV, but 
ATLAS had to add more Run 2 data to some of the modes.\cite{Aaboud:2018urx} 
The process is just radiating $H$ off the top in the QCD production of $t\bar t$.
But why is there no hoopla?
From my Flavor background, the direct measurement of 
the Top Yukawa coupling, $\lambda_t$, and finding 
consistency with the expected SM value of $\sim 1$, is a true landmark event.
After all,  $\lambda_t$ is {\bf\it the Mother of most SM loop effects}, 
be it Flavor/B-Physics, or effective $ggH$, $\gamma\gamma H$ couplings.
Alas, it went also largely ignored by flavor folks at the 
FPCP2018 conference held July 2018 in Hyderabad, India.

Together with observation of $H \to \tau^+\tau^-$ and $H \to b\bar b$ (see Table 1),
all consistent with SM expectations, we have now measured
third generation Yukawa couplings {\it directly}, confirming $\lambda_f = \sqrt{2}m_f/v$,
where $v$ is the vacuum expectation value (v.e.v.) of the SM Higgs field.
But people seem to want signs for New Physics so badly, observing SM-like 
$t\bar tH$ (and $H\tau^+\tau^-$ and $Hb\bar b$) coupling is not good enough.

\subsection{Flavor, where the Anomalies/hoopla are!}

There is no doubt that all current ``anomalies'' are in the flavor sector.~\cite{Passaleva,Eklund}
We have, in order of the first announcements:\cite{Chang:2017wpl} 
\begin{itemize}
 \item $R_{D^{(*)}}$ anomaly:
   the ratio of $B \to D^{(*)}\tau\nu$ rate with $B \to D^{(*)}\mu\nu$;
 \item $P_5'$ anomaly: the angular variable in some 
   $q^2 = m_{\ell\ell}^2$ bin(s) of $B \to K^*\mu^+\mu^-$; 
 \item $R_{K^{(*)}}$ anomaly:
   the ratio of $B \to K^{(*)}\mu^+\mu^-$ rate with $B \to K^{(*)}e^+e^-$.
\end{itemize}

\subsubsection{High $p_T$ Response to Flavor Anomalies}

After much hoopla from theorists in the past 5--6 years, 
two popular NP pictures have caught the attention of 
high-$p_T$ experiments in 2018: $Z'$, and leptoquark (LQ).

\begin{figure}[t]
\hskip0.1cm\vskip0.2cm
\begin{minipage}{0.32\linewidth}
\centerline{\includegraphics[width=1.01\linewidth]{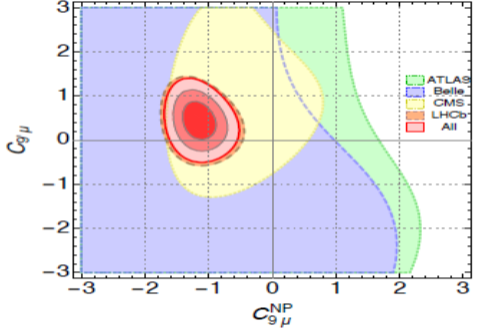}}
\end{minipage}
\hskip-0.1cm
\begin{minipage}{0.32\linewidth}
\centerline{\includegraphics[width=1.06\linewidth]{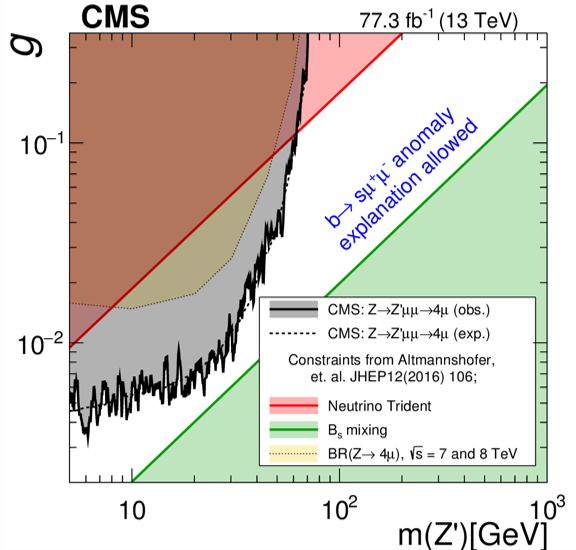}}
\end{minipage}
\begin{minipage}{0.32\linewidth}
\centerline{\includegraphics[width=1.13\linewidth]{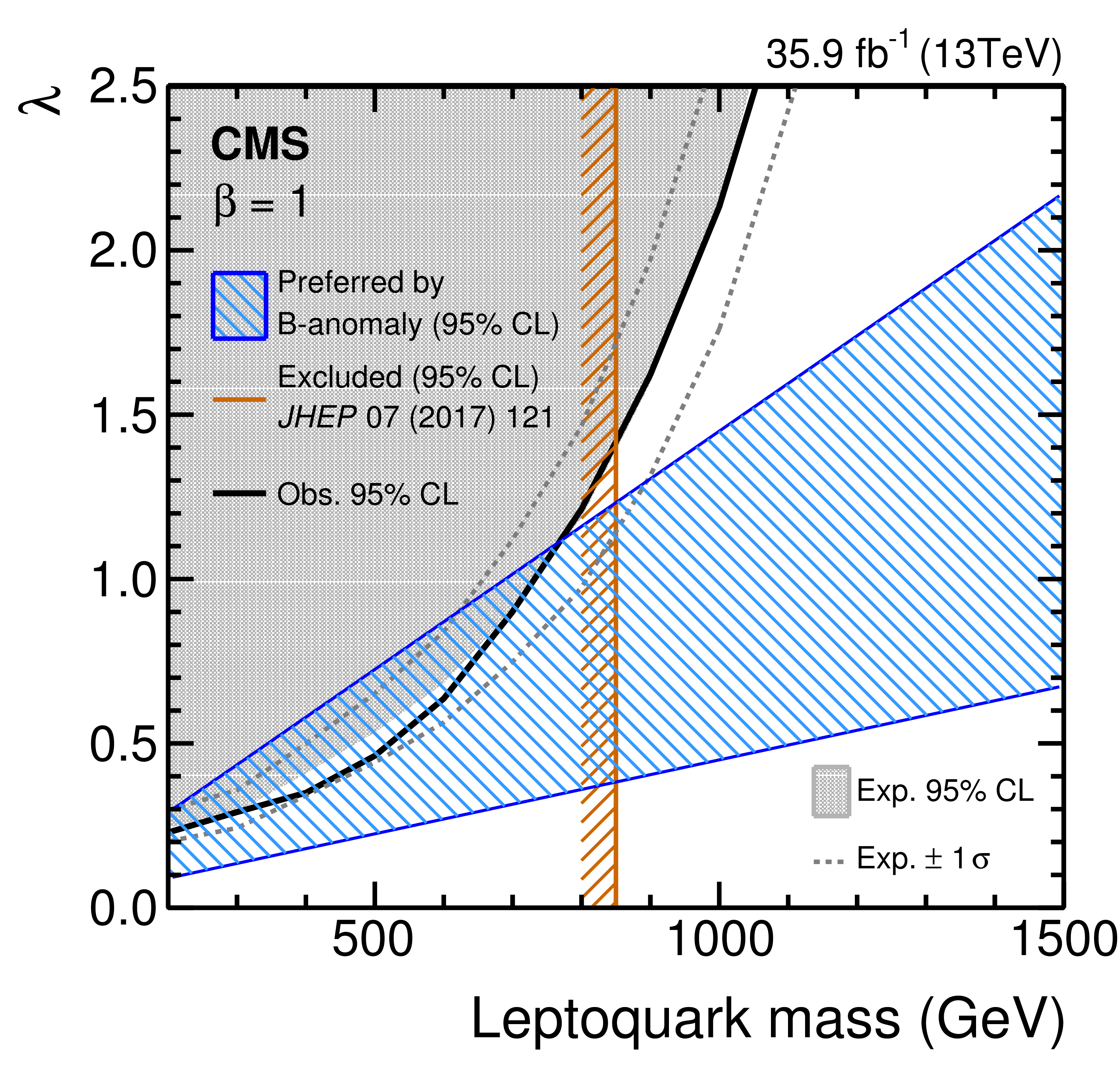}}
\end{minipage}
\caption{
 [left] Fit \cite{Capdevila:2017bsm}to $R_{K^{(*)}}$ and $P_5'$;
 CMS search for [center] $Z'$,\cite{Sirunyan:2018nnz} and
 [right] $\tau b$ leptoquark.\cite{Sirunyan:2018jdk}
}
\label{fig:anomalies}
\end{figure}

But let us first stress the impressive fit (see Fig.~\ref{fig:anomalies}[left]), 
that $R_{K^{(*)}}$ and $P_5'$ can be accounted for  
by a shift,\cite{Capdevila:2017bsm} $\Delta C_9 \simeq -1$, in the $C_9$ Wilson coefficient,
hence of similar strength to SM,
where we cite only a sample reference (similarly below).
A common interpretation is a $Z'$ mediating the $b \to s\ell^+\ell^-$ decays,
which could be\cite{Altmannshofer:2014cfa} 
the gauge boson of $L_\mu - L_\tau$ symmetry.
CMS took notice of this\cite{Butler} and made a specific search 
for the $L_\mu - L_\tau$ gauge boson\cite{Altmannshofer:2016jzy}
via on-shell $Z \to \mu^+\mu^-Z' \to \mu^+\mu^-\mu^+\mu^-$
using $\sim 77$ fb$^{-1}$ at 13 TeV,\cite{Sirunyan:2018nnz} 
which is illustrated in Fig.~\ref{fig:anomalies}[center].
Note that searching in $Z$ decay limits sensitivity to relatively light $Z'$.

When all things are considered, LQ is favored for the $R_{D^{(*)}}$ anomaly,
which can in principle also account for $R_{K^{(*)}}$.
High $p_T$ again took notice, and CMS made it a highlight result\cite{Rahatlou}
at the ICHEP2018 summer conference held in Seoul, pursuing\cite{Sirunyan:2018jdk} 
singly produced $\tau b$-LQ via $bg \to \tau^- {\rm LQ} \to \tau^-\tau^+ b$,
to probe the parameter space (see Fig.~\ref{fig:anomalies}[right])
of some combined model projection.\cite{Buttazzo:2017ixm} 
The vertical line in the plot at $\sim 850$ GeV is from 
an earlier LQ-pair production search,\cite{Sirunyan:2017yrk}
which has just been updated by CMS to 1.02 TeV,~\cite{Sirunyan:2018vhk}
to be compatible with the 13 TeV dataset used for the single-LQ production search.

While I would emphatically state my bias {\it against} LQs (``Why LQs Now!?''),
the significance of these two searches is that high-$p_T$ experiments
are now paying attention to flavor anomalies.

We have already mentioned that there is no lack of theory activities 
on the flavor anomalies, and they are indeed too numerous to mention.
The plenary speaker at {\it Rencontres du Vietnam}, Gino Isidori,\cite{Isidori}
also chose to present a perspective (that bypassed $P_5^\prime$).
It was first emphasized that the stated anomalies,
``{\it IF} taken together, \ldots is probably the largest
 `coherent' set of NP effects in present data \ldots''
And that, ``What is particularly interesting, is that these anomalies
are challenging an assumption -- LUV -- that we gave for granted 
for many years (without much good theoretical reasons \ldots).''
With LUV standing for Lepton Universality Violation, 
we certainly concur with the last statement, as well as 
to keep an eye on $\tau \to \mu$ LUV processes.\footnote{
 It should be mentioned that there was a hint\cite{Khachatryan:2015kon}
 by CMS Run 1 data for sizable $h \to \tau\mu$ decay,
 at $\sim 1\%$ level with 2.4$\sigma$ significance. 
 Together with the theory suggestion\cite{Harnik:2012pb} 
 that prompted the $h \to \tau\mu$ search at LHC,
 interest in $\tau \to \mu$ LUV processes such as $\tau \to \mu\gamma$, $3\mu$ grew.
 Unfortunately, with $\sim 36$ fb$^{-1}$ at 13 TeV,\cite{Sirunyan:2017xzt}
 CMS ruled out the previous hint, setting the stringent bound of
 ${\cal B}(h \to \tau\mu) < 0.25\%$.
 But $\tau \to \mu\gamma$, $3\mu$ and other modes should still be watched.
}
We have already cited the simplified LQ model work\cite{Buttazzo:2017ixm}
that CMS set forth to probe, as discussed above.
But Gino went on to sophisticated (UV) model building,\cite{Isidori}
 such as ``PS$^3$'',\cite{Bordone:2017bld} i.e. having 3 copies of Pati-Salam symmetry
to accommodate the flavor (mass and mixing) hierarchies,
which is becoming a bit much for our taste.
It does illustrate that there are no convincing models out there
to cover all the flavor anomalies.

\subsubsection{Experimental Caution/Reminder on Flavor Anomalies}

Inasmuch as they are our current {\it Best Hopes} for BSM indications,
but keeping in mind that Physics is an experimental science,
let me put forth the grains of salt I have regarding these flavor anomalies 
from the experimental viewpoint.\footnote{
 Well, call it the advice from an ``Experimentheorist''.
}

\begin{figure}[t]
\vskip-0.3cm\centerline{
\includegraphics[width=5.9cm]{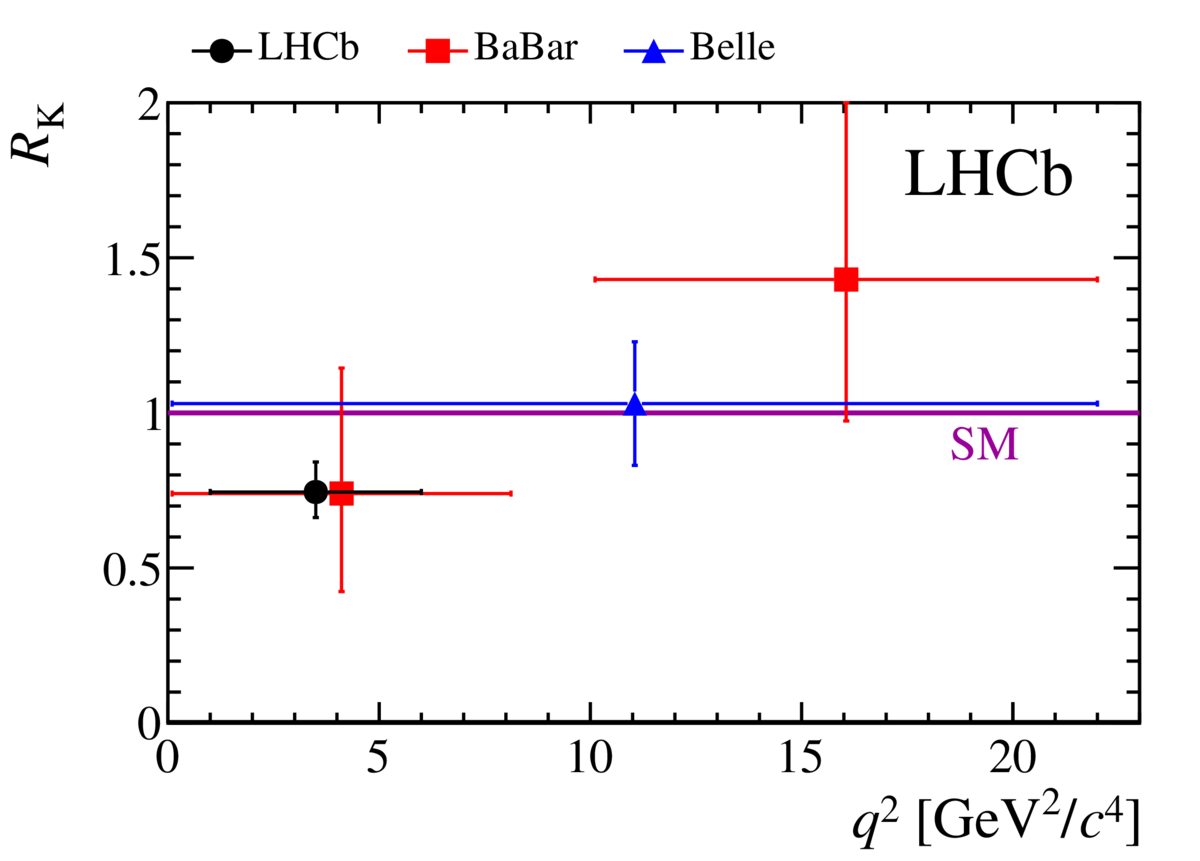}
\includegraphics[width=6cm]{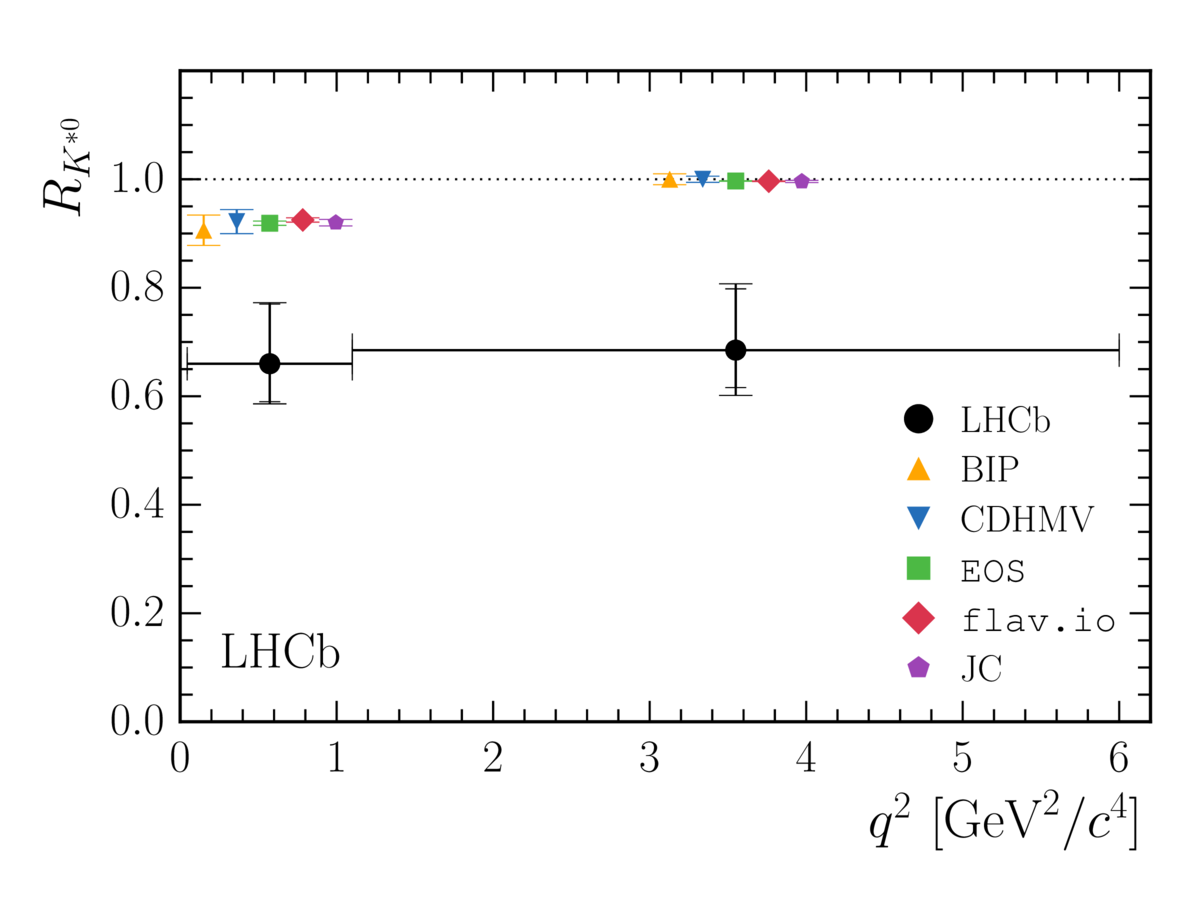}
}
\caption[]{
Measured $R_K$ and $R_{K^*}$ values by LHCb\cite{Aaij:2014ora,Aaij:2017vbb}
compared with SM expectation and with other experiments.
[$R_K$ figure from https://lhcbproject.web.cern.ch,
 $R_{K^*}$ figure from Ref.~\citen{Aaij:2017vbb}]
}
\label{fig:RK-RKst}
\end{figure}

\vskip0.2cm
\noindent \underline{LUV:  $R_K$, $R_{K^*}$ Anomalies}

Must Love the LUV: 
The $R_{K^{(*)}}$ ratios which test Lepton Universality are theoretically clean!
But, after the original $R_K$ result\cite{Aaij:2014ora} of 2014 that showed $-2.6\sigma$ 
discrepancy with SM for $q^2$ in $1.0-6.0$ GeV$^2$, 
it took 3 years for LHCb to put forth\cite{Aaij:2017vbb}
 the $R_{K^*}$ measurement (see Fig.~\ref{fig:RK-RKst}). 
The two $R_{K^*}$ bins are for $q^2$ in $0.045 - 1.1$ GeV$^2$
and $1.1-6.0$ GeV$^2$, showing $-2.2\sigma$ and $-2.4\sigma$ 
downward shifts, respectively.\footnote{
 In fact, the downward shift from 1.0 is the largest for the lower $q^2$ bin,
 but its SM expectation is lower because of dimuon threshold effect.
}
 Looks brilliant.
However, why change from 1.0 to 1.1 GeV$^2$? 
It is to cover the $\phi$ meson: the lower $R_{K^*}$ bin is dominated by the photon.
But the photon coupling to $e$ and $\mu$ is {\it certainly Universal}.
So, at EPSHEP2017 summer conference held in Venice, I had cautioned relatively loudly
with something like, ``I will bet that LHCb measured the lower $R_{K^*}$ bin
as a sanity check, since it is expected to be consistent with SM."
If SM was confirmed in the lower bin, LHCb would have proclaimed victory, and then  
combine the two corresponding (in $q^2$) $R_K$ and $R_{K^*}$ bins.
Given the common drop for all three measurements, I stress that 
one ought to follow Occam's advice and keep in mind the ``simpler'' explanation: 
rather than NP, could it be some common systematics, perhaps traced to 
normalizing\cite{Aaij:2017vbb} on $B \to K^{(*)}J/\psi\, (\to \mu^+\mu^-,\,e^+e^-)$?

\vskip0.2cm
\noindent \underline{Is $P'_5$ Anomaly Real?}

Published in 2013 using 1 fb$^{-1}$ data,\cite{Aaij:2013qta} 
the $P'_5$ anomaly was LHCb's first, finding 3.7$\sigma$ discrepancy with SM 
in a particular $q^2$ bin of the angular variable $P_5'$. 
This already lead to many theory papers toting $\Delta C_9 \sim -1$ in 2013,
but questions also arose whether this was a fluctuation among many measurables,
or possibly due to the $c\bar c$ threshold, i.e. hadronic, effects.
But perhaps more symptomatic was when LHCb announced 
the 3 fb$^{-1}$ result at Moriond 2015, 
significance of the discrepancy dropped slightly\cite{Aaij:2015oid}  to 3.4$\sigma$,
despite the increase in statistics allowed splitting the bin into two.
That is, while errors improved, the central values of both bins moved closer to SM.
Statistically speaking, there is nothing ``wrong'' with this, 
but when I first learned about it, I quipped that this is
 ``Not Good, Not Bad \ldots \underline{Not Too Good}''.
If the original measurement was close to ``Truth'', one would
expect the significance to improve somewhat when data tripled.
It also makes the $c\bar c$ threshold issue more troubling.\cite{Ciuchini:2015qxb}
Adding on top of this was the CMS announcement in 2017,
based on 8 TeV data, where the values in the two bins
are consistent with SM.~\cite{Sirunyan:2017dhj} 
Note that Belle has one broad bin that is consistent with LHCb but with less resolution,
while ATLAS lacks a second bin, even though the first bin shows wide deviation.

The issue can only be resolved by more data,
preferably by multiple experiments.

\begin{figure}
\centerline{\includegraphics[width=0.65\linewidth]{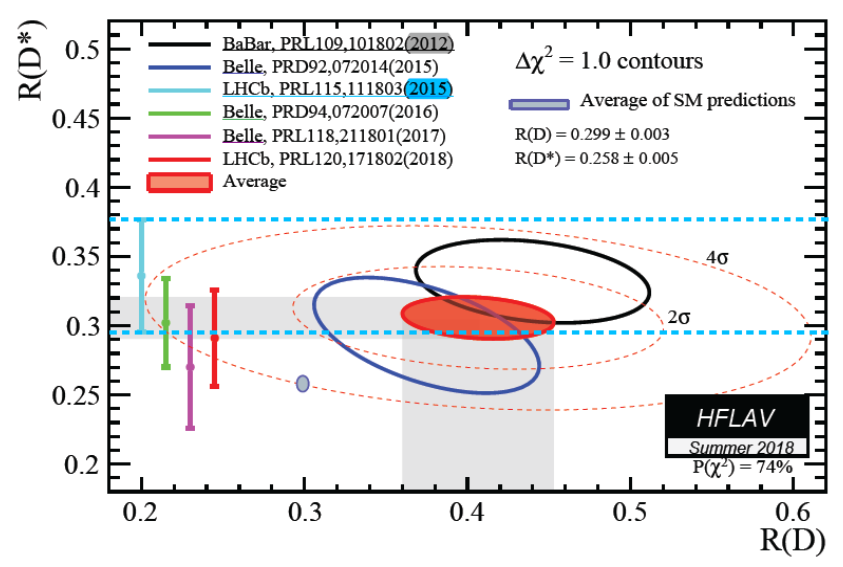}}
\caption[]{
 Summer 2018 HFLAV\cite{HFLAV} combination of $R_D$ vs $R_{D^*}$. 
 SM is the very small gray ellipse.
}
\label{fig:RD-RDst}
\end{figure}

\vskip0.2cm
\noindent \underline{Is $R_{D^{(*)}}$ Anomaly Real?}

Fig.~\ref{fig:RD-RDst} from\footnote{ 
 The acronym was changed from HFAG since Moriond 2017.
} 
Heavy Flavor Averaging Group (HFLAV) gives\cite{HFLAV} 
the history and evolution of  $R_D$--$R_{D^*}$ measurement, 
where discrepancy of world average (central ``red'' ellipse) and SM (very small gray ellipse)
 is at 3.7$\sigma$.
The two larger ellipses are the measurements by BaBar and Belle
in 2012 and 2015, respectively, with Belle disagreeing
neither with BaBar, nor with SM. The BaBar result\cite{Lees:2012xj}
of course already caught interest of theorists.
But the surprise announcement\cite{Aaij:2015yra} by LHCb in 2015 
of its prowess to measure $R_{D^*}$, where the horizontal
dashed-line band concurred with BaBar, 
turned it into a sensation for the theory world.
What we wish to caution is in regards the two most recent results:
the $R_{D^*}$ measurement by LHCb,\cite{Aaij:2017uff}  and
the $R_{D^*}$  and $\tau$-polarization measurement by Belle,\cite{Hirose:2017dxl} 
both published in 2018 and utilize $\tau \to$ 3-prong
 (rather than $\tau \to \mu\nu\nu$ for all previous studies),
and are in full agreement with SM. 
Could there be some common systematics in taking the ratio of 
$B \to D^{(*)}\tau\nu$ to $B \to D^{(*)}\mu\nu$ with $\tau \to \mu\nu\nu$ decay?

The issue clearly can only be resolved by more data, plus dilig/vigilence.

\subsection{Whither Extra Yukawas?}

While we eagerly await the LHCb update of any of the
flavor anomalies with Run~2 data (indeed, the silence has been deafening), 
let me finally offer some personal perspective,
 on {\it Extra Yukawa Couplings}, which I deem as 
{\it\bf the most likely, next, New Physics}.

All CPV measured in the laboratory so far are accountable by the Kobayashi-Maskawa phase, 
which is rooted in Yukawa couplings, as are CKM matrix elements.
This naturally begs the question: 

\centerline{\underline{\it Are there Extra Yukawas?}}

\vskip0.15cm\noindent
After all, the Jarlskog invariant is way too small in SM for the
disappearance of Antimatter from the Universe.\cite{Holdom:2009rf}
With a second Higgs doublet, one would naturally have
a second set of Yukawa couplings.
Alas, extra Yukawas in usual 2HDM were killed by 
the Natural Flavor Conservation (NFC) condition\cite{Glashow:1976nt}
of Glashow and Weinberg (usually realized by a $Z_2$ symmetry),
for fear of FCNH couplings.

\subsubsection {Nothing Natural about NFC}

But with one Higgs doublet confirmed by the $h(125)$ discovery,
a 2nd doublet is now highly plausible, and the NFC condition should be reexamined.
We advocated recently the $tch$ coupling,~\cite{Chen:2013qta}
\begin{equation}
\rho_{tc} \cos(\beta - \alpha)\, \bar tch,
\end{equation}
where $\rho_{tc}$ belongs to the least constrained extra Yukawa couplings
in a 2HDM without $Z_2$: $\rho_{cc}$, $\rho_{ct}$, $\rho_{tc}$, $\rho_{tt}$.
The $\rho_{ct}$ coupling is actually constrained 
by $B$ physics to be rather small,\cite{Altunkaynak:2015twa}
and I now think $\rho_{ii}$ should be ${\cal O}(\lambda_i)$,
i.e. $\rho_{cc}$ should be suppressed by $m_c/v \lesssim 0.005$,
but $\rho_{tc},\; \rho_{tt}$ ought to be ${\cal O}(\lambda_t) \sim 1$.

Eq.~(1) improves on the Cheng--Sher trickle-down argument\cite{Cheng:1987rs}
(which I coined the ``type III 2HDM'' name) 
when I first advocated $t \to ch$ decay a long time ago:\cite{Hou:1991un} 
the $tch$ coupling is modulated by $\cos(\beta - \alpha)$,
the $h$--$H$ mixing angle, where $H$ is the exotic $CP$-even neutral Higgs
of the extra Higgs doublet. In this sense, if $\cos(\beta - \alpha)$ is small,
which is reflected in the apparent alignment phenomenon
that $h$ appears rather close to SM Higgs,\cite{Khachatryan:2016vau} 
the non-observation of $t \to ch$ so far (see Sec.~1.6 and Fig.~\ref{fig:tCNC}) 
need not imply a small $\rho_{tc}$ as yet.
So, in place of NFC, the trickle-down flavor pattern, alignment, 
and $1/m_H$ suppression can work together to
suppress effects of off-diagonal $\rho_{ij}^{(f)}$ Yukawa couplings,
 where $f = u,\; d,\; \ell$.
Glashow and Weinberg need not have invoked NFC over 40 years ago 
to protect against FCNH.

Note that, without a $Z_2$ symmetry, one cannot actually
distinguish the $\Phi_1$ and $\Phi_2$ doublets,
hence $v_1$ and $v_2$, and thus $\tan\beta = v_1/v_2$ is ill-defined.
We therefore replace $\cos(\beta - \alpha)$ in Eq.~(1) by $\cos\gamma$,
the notation which was first used in Ref.~\citen{Hou:2017vvp}.
We note that this reference provides a one-loop protection mechanism 
for alignment: with the right sign, $\rho_{tt} \sim {\cal O}(\lambda_t) \sim 1$
can restore $|\sin\gamma| \to 1$ from sizable bosonic loop corrections 
arising from ${\cal O}(1)$ Higgs quartic couplings.

\subsubsection{Electroweak Baryogenesis}

With ${\cal O}(1)$ Higgs quartic couplings able to give rise to\cite{Kanemura:2004ch} 
1st order electroweak phase transition (EWPT),
it was noted recently\cite{Fuyuto:2017ewj} that 
$\rho_{tt} \sim {\cal O}(1)$ can come hand in hand 
to provide a rather robust mechanism for electroweak baryogenesis (EWBG).
The extra top Yukawa coupling $\rho_{tt}$ is naturally complex,
and as the 33 element of the combination of the two Yukawa matrices
that is orthogonal to the one that gave the up-type mass matrix, 
it is also naturally ${\cal O}(1)$. 
A study of the top scattering at the expanding bubble wall 
(from 1st order EWPT) gives the leading CPV source as\cite{Fuyuto:2017ewj}
\begin{equation}
\lambda_t\,{\rm Im}\rho_{tt},
\end{equation}
which suffers no suppression factors, compared with the
multiple suppression by small-Yukawa-couplings for the Jarlskog invariant in SM.

\begin{figure}[t]
\centerline{
\includegraphics[width=6.7cm]{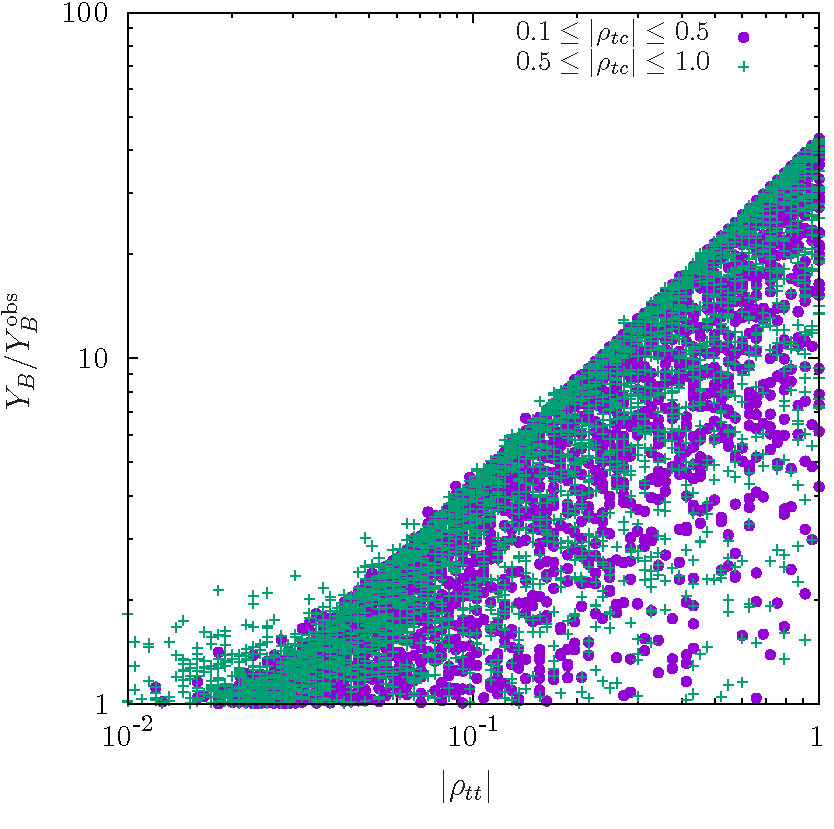}
}
\caption[]{
Scatter plot of $Y_B/Y_B^{\rm obs}$ vs $|\rho_{tt}|$,
scanning through $|\rho_{tc}|$ and the 
phases of $\rho_{tt}$ and $\rho_{tc}$.\cite{Fuyuto:2017ewj}
}
\label{fig:EWBG}
\end{figure}

Scanning over $|\rho_{tc}|$ and the phases of $\rho_{tt}$ and $\rho_{tc}$,
we produce the scatter plot (Fig.~\ref{fig:EWBG}) of $Y_B/Y_B^{\rm obs}$,
 the ratio of baryon production with respect to what is observed (e.g. by Planck), 
versus $|\rho_{tt}|$.
With a ``margin'' of almost two orders of magnitude, 
the mechanism is indeed rather robust.
We note also that, in case $\rho_{tt}$ is accidentally small, 
from the (green) crosses that correspond to larger $|\rho_{tc}|$ values,
a second backup mechanism can come from sizable $\rho_{tc}$
 with near maximal CPV phase.
In making this plot, $\rho_{tc}$ and $\rho_{tt}$ are 
checked against\cite{Altunkaynak:2015twa} $B_d$, $B_s$ 
mixings and $b\to s\gamma$ constraints.
Also, the exotic Higgs bosons were held degenerate at
$m_H = m_A = m_{H^+} = 500$ GeV for simplicity.
On one hand, the parameter space is even broader.
On the other hand, the sub-TeV values make the scenario
even more attractive.

\subsubsection{Extra Yukawa as Experimental Issue}

The existence of Extra Yukawas is actually an experimental issue.

When we discovered $m_h < m_t$, nothing stops the experimentalist,
young or old, to search for $t\to ch$, because 
it can, and must (PDG entry!), be done.
However, unlike most New Physics arising from some high scale, 
this is a dimension-4 term, i.e. a regular Lagrangian term in 
the form of a Yukawa coupling, hence immediately implies 
the possible existence of an extra doublet that can give rise to
 extra Yukawas. 

Thus, logically, ever since the Higgs boson discovery,
we have been probing for Extra Yukawa couplings via $t\to ch$ 
(and $h \to \tau\mu$) search.
If discovery is made any time soon, it would have to be a tree-level effect,
rather than the ultra-suppressed loop-induced effect in SM.
This explains my advocacy, or fondness, of Fig.~\ref{fig:tCNC}.

The logical extension, then, is to treat all extra Yukawas, 
$\rho_{ij}^{(f)}$ ($f = u$, $d$, $\ell$), as experimental issues.

\subsubsection{Curious: Alignment as Emergent}

From the proximity of $h$ to the SM-Higgs,
we have already mentioned that $|\sin\gamma|$ is rather close to 1,
hence $|\cos\gamma|$, the mixing parameter between $h$ and $H$,
is rather small. This is a little puzzling if $m_H$ is sub-TeV
(rather than decoupled at several TeV).
What may be worse, it superficially runs against ${\cal O}(1)$
Higgs quartic couplings, which can be felt as generally inducing
sizable $|\cos\gamma|$.

\begin{figure}[t]
\centerline{
\includegraphics[width=7cm]{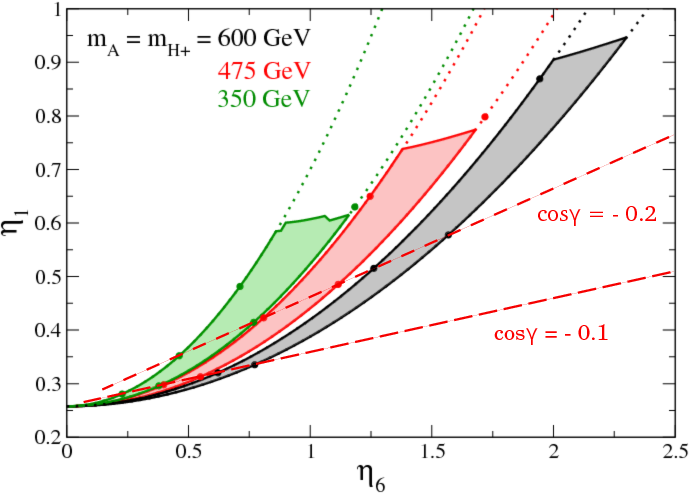}
}
\caption[]{
Allowed parameter space\cite{Hou:2017hiw} in Higgs quartic couplings $\eta_1$ and $\eta_6$
 corresponding to alignment (small $\cos\gamma$).
}
\label{fig:Align}
\end{figure}

With this fuzzy backdrop, it was found\cite{Hou:2017hiw} that, 
curiously, there is considerable parameter space for alignment.
As Higgs quartics are numerous in number, and also to face
electroweak precision measurement constraint, we illustrate\cite{Hou:2017hiw}
in Fig.~\ref{fig:Align} the allowed parameter space
in $\eta_1$ (controls $m_h$) vs $\eta_6$ (controls $h$--$H$ mixing),\footnote{
 For 2HDM without $Z_2$, there are the extra $\eta_6$ and $\eta_7$ couplings,
 but only one extra parameter, as the usual ``soft breaking'' parameter
 gets absorbed into $\eta_6$ by minimization condition.\cite{Hou:2017hiw}
}
where custodial SU(2) symmetry is assumed such that $m_A = m_{H^+}$
so the $T$ parameter constraint is easier to accommodate
(which cuts off the horn shaped regions for given  $m_A = m_{H^+}$).
To the lower left, the tip of the horn, one has extreme alignment
as $\eta_6 \to 0$ and $\eta_1 \to m_h^2/v^2$.
But there is much solution space for $\eta_6$, even $\eta_1$,
being ${\cal O}(1)$.
The true solution space is even larger if one just impose 
the EW precision constraints, but it becomes harder to plot.
Thus, the alignment phenomenon is not difficult to accommodate
for sub-TeV exotic Higgs bosons.
Away from extreme alignment, or alignment limit,
one has\cite{Hou:2017hiw,Bechtle:2016kui} the classic looking formula,
\begin{equation}
\cos\gamma \cong \frac{-\eta_6 v^2}{m_H^2 - m_h^2},
\end{equation}
where one can easily achieve $|\cos\gamma| < 1/4$ for
sufficiently large $m_H^2$--$m_h^2$ splitting but sizable $\eta_6$. 
In fact a finite $\eta_6$ mixing parameter can help by 
level repulsion, as can also be seen from Fig.~\ref{fig:Align}.
We note that the diagonal contribution to $m_H^2$ is fed by 
the Higgs quartics $\eta_3$, $\eta_4$, $\eta_5$
and the decoupling mass $\mu_{22}^2$,\cite{Hou:2017hiw}
and could easily be larger than $m_h^2$, which has only the
diagonal contribution coming from $\eta_1$.

\subsubsection{Upshot: Top-Higgs,  to ``Flavored'' Higgs, to  SM2?}

To summarize, two observations 
\begin{enumerate}
\item ${\cal O}(1)$ Higgs quartic couplings plus ${\cal O}(1)$ complex $\rho_{tt}$ (or $\rho_{tc}$)
   give remarkably efficient EWBG;\cite{Fuyuto:2017ewj}
\item ${\cal O}(1)$ Higgs quartic couplings, needed for 1st order EWPT,
   can readily support\cite{Hou:2017hiw} approximate alignment! 
\end{enumerate}
convince me that ``Extra Yukawas via 2HDM without $Z_2$''
may well be the next New Physics.
To reduce the mouthful of words by removing redundancy, 
we recently dubbed\cite{Chang:2017wpl} it
``SM2'', i.e. SM with a 2nd Higgs doublet
 --- no added assumptions and just {\it let Nature speak}.
This SM2 can be probed at LHC via $cg \to tH^0,\; tA^0 \to t\bar tc$ (same-sign top), 
$tt\bar t$ (triple-top) signatures,~\cite{Kohda:2017fkn} and may impact on 
$B^+ \to \mu^+\nu$\cite{Sibidanov:2017vph} and 
electron electric dipole moment (e-EDM),\cite{Andreev:2018ayy}
to name just a few processes.

If I may raise any caution, I must say that we know very little, experimentally, 
about the Higgs potential, or Higgs sector self-couplings. 
But the flavor side, i.e. extra Yukawa couplings, stand on firmer ground, 
as we already know that a plethora of Yukawa couplings exists in SM,
which Nature has expressed the nontrivial mass and mixing hierarchies.

\section{Run 2 Era Looks Bright, and Flavorful}

The LHC Run 2 has come to an end. However,
with all data now at hand, in a sense the Run 2 Era has just begun.
Judging from Run 1 and LS1, 
many good results based on full Run 2 data would appear
during LS2, extending actually into 2021, the first year of Run 3 start,
and even beyond.
Thus, I will continue to call 2019--2021,
the next three years, as ``Run 2 Era''.

The ATLAS and CMS experiments have collected more than 
$5\times$ the data of Run~1, and at higher energy.
The LHCb experiment has collected more than
$2\times$ the data of Run 1, and at higher energy.
This already makes the Outlook Bright, brighter than anticipations for Run 3 Era,
where only a doubling of Run 2 data is expected,
with little change in energy.
Furthermore, the $B$ physics program of Belle~II would
commence in Spring 2019, ushering in a new decade,\footnote{
Partially in response, LHCb is rebuilding its detector, Upgrade 1,
during LS2.
LHCb has also announced its formidable Upgrade 2 plan.\cite{Bediaga:2018lhg}
} 
where dark photon, $R_D$--$R_{D^*}$ anomaly, $B^+ \to \mu^+\nu$
and other measurements, clarifications and searches could~\cite{Urquijo} 
fall into our ``Run 2 Era'' (i.e. by 2021).
See the Belle II Physics Book for more discussion.~\cite{Kou:2018nap}

Besides Belle II and the LHCb-driven flavor anomalies, 
there are many experiments related to flavor that are ongoing, 
such as 
\begin{itemlist}
\item First result by NA62 at CERN on $K^+ \to \pi^+\nu\nu$ 
 search,~\cite{Aliberti,CortinaGil:2018fkc} targeting
 SM measurement eventually;
\item The recent order of magnitude improvement~\cite{Ahn:2018mvc} of 
 $K_L \to \pi^0\nu\nu$ bound by KOTO at KEK, and search for exotic
 $K_L \to \pi^0 X^0$,~\cite{Fuyuto:2014cya} where 
 $X^0$ is a ``dark'' object with mass around $\pi^0$,
 with more data at hand plus continued running;
\item The new muon g--2 experiment at Fermilab,~\cite{Tran,g-2}
 where we may hear first results as early as 2019,
 then the second installment in 2020 (where data taking will end),
 and the definitive result might be delivered by 2021 (!);
\item Having reached the impressive\cite{TheMEG:2016wtm}
 ${\cal B}(\mu \to e\gamma) < 4.2 \times 10^{-13}$ at 90\% C.L.,
 MEG~II at PSI aims\cite{Baldini:2018nnn} to improve by almost
 another order of magnitude in the next few years;
 it remains to be seen whether the two new $\mu$ to $e$ conversion experiments,
 COMET\cite{COMET} at KEK vs Mu2e\cite{Mu2e} at Fermilab,
 would produce results within this time frame,
 but the competition is strong.
\end{itemlist}

\noindent\underline{In conclusion}, our Run 2 Era of 2019--2021 is not only {\it Bright},
but would be {\it Flavorful}, and let's hope it would be Wonderful.
While we have put forth grains of salt for each of the flavor anomalies,
if just one of them holds true, we would have lucked out.
And with so many directions that we have mentioned,
the next three years, the extended Run 2 Era,
may just be {\it Golden}! And it may well extend into a decade.
We advocate a most likely, next New Physics, 
to reveal itself in the next 3--5 years,
would arise from {\it Extra Yukawa Couplings}.

\section*{Acknowledgments}

I thank Meenakshi Narain and the Program Committee of
the 25th Anniversary Conference of the Rencontres du Vietnam 
for the invitation to speak, and Jean Tran Than Van and the LOC for hospitality.
I am grateful to all the plenary speakers who helped realize the HEP summary talk. 
This work is supported by grant MOST 106-2112-M-002-015-MY3 of Taiwan.


\end{document}